\documentclass{llncs}
\usepackage{graphicx}
\usepackage{caption}
\usepackage{amsmath}
\usepackage{enumerate}
\usepackage{subfig}
\usepackage{float}
\usepackage{amsfonts}
\usepackage{hyperref}
\usepackage[all]{hypcap}

\begin{document}
\frontmatter          
\pagestyle{headings}  
\addtocmark{Dynamics of Random Boolean Threshold Networks} 

\title{Boolean Threshold Networks: Virtues and Limitations for Biological Modeling}
\titlerunning{Dynamics of Random Boolean Threshold Networks}  
%
\author{Jorge G.T. Za\~nudo, Maximino Aldana and Gustavo Mart\'inez-Mekler}
\authorrunning{Za\~nudo et al.}   
%
\tocauthor{J.G.T. Za\~nudo, M. Aldana, G. Mart\'inez-Mekler}
\institute{Instituto de Ciencias F\'isicas, Universidad Nacional Aut\'onoma de M\'exico.\\
Avenida Universidad s/n, Colonia Chamilpa, Cuernavaca, Morelos, M\'exico. C\'odigo Postal 62210.\\
\email{max@fis.unam.mx},\\ WWW home page:
\texttt{http://www.fis.unam.mx/\homedir max/}}

\maketitle              

\begin{abstract}
Boolean threshold networks have recently been proposed as useful tools to model the dynamics of genetic regulatory networks, and have been successfully applied to describe the cell cycles of \textit{S. cerevisiae} and \textit{S. pombe}. Threshold networks assume that gene regulation processes are additive. This, however, contrasts with the mechanism proposed by S. Kauffman in which each of the logic functions must be carefully constructed to accurately take into account the combinatorial nature of gene regulation. While Kauffman Boolean networks have been extensively studied and proved to have the necessary properties required for modeling the fundamental characteristics of genetic regulatory networks, not much is known about the essential properties of  threshold networks. Here we study the dynamical properties of these networks with different connectivities, activator-repressor proportions, activator-repressor strengths and different thresholds. Special attention is paid to the way in which the threshold value affects the dynamical regime in which the network operates and the structure of the attractor landscape. We find that only for a very restricted set of parameters, these networks show dynamical properties consistent with what is observed in biological systems. The virtues of these properties and the possible problems related with the restrictions are discussed and related to earlier work that uses these kind of models.
\end{abstract}

\section{Introduction} \label{sec:1}

The analysis of the dynamics of genetic regulatory networks in living organisms is a complicated task and a central challenge in current research for a complete understanding of complex biological systems. Historically, the dynamical behaviour of the biochemical elements in small genetic circuits has been accurately described using differential equations, which capture the underlying reaction-diffusion kinetics that take place in these systems \cite{1,2,3}. However, this approach faces important difficulties for the modeling of large genetic networks, being the main difficulty that these mathematical models may involve a very large amount of parameters. This is a serious problem both practically and theoretically. Practically since these parameters may be largely unknown for many systems, and theoretically since such a detailed description may obscure the essential properties of the regulatory processes in the systems under consideration\cite{4,Reka-no-kinetic-detail}. Because of this, Boolean networks have recently been increasingly used as the best first approach for the modeling and understanding of the essential properties of real regulatory systems that incorporate large amounts of data \cite{Reka-discrete-10,Reka-boolean-04}.

Boolean networks have been extensively studied for decades \cite{7}, and were introduced for the modeling of large regulatory systems by S. Kauffman as a first attempt to understand the general dynamical properties of the gene regulation and cell differentiation processes \cite{5,6}. However, it was only recently that the necessary information to test them on real biological genetic networks has been available. Examples are models of the genetic network of flower development in \emph{Arabidopsis thaliana} \cite{8,9}, the regulatory network determining embryonic segmentation in \emph{Drosophila melanogaster} \cite{10}, the network controlling the differentiation process in Th cells \cite{40}, the cell cycle networks of \emph{Saccharomyces cerevisiae} \cite{11} and \emph{Saccharomyces pombe} \cite{12}, among others. One of the advantages of the Boolean approach is that it is not necessary to know the kinetic details of the interactions (e.g. promoter affinities, degradation constants, translation rates, etc.). Rather, only the logic of the regulatory interactions is needed, such as the specific activatory or inhibitory nature of the genetic regulations \cite{4,Reka-no-kinetic-detail}. By incorporating this information, available nowadays from high-throughput experiments, into the Boolean approach, it has been possible to predict the temporal sequence of gene activities as well as the stable and periodic patterns of gene expression in wild type and in many mutants of the organisms mentioned before.

There are, however, important differences in the way in which Boolean models have been implemented by different groups, and it is not clear whether or not these different implementations would yield equivalent results. The general formulation of the Boolean network model is the following.  We assume that the network is represented by a set of $N$ Boolean variables (or genes) $\left\{\sigma_1, \sigma_2, \ldots, \sigma_N \right\}$, each of which can be in two different states $\sigma_i=1$ (active) and $\sigma_i=0$ (inactive). The state of each gene $\sigma_i$ is controlled by $k_i$ other genes of the network, $\{ \sigma_{i_1}, \ldots, \sigma_{i_{k_i}} \}$, which we will refer to as the \emph{regulators} or the \emph{inputs} of $\sigma_i$. The number $k_i$ of regulators of each gene depends on the topology of the network in such a way that the probability for a randomly selected node to have $k$ regulators is given by the probability distribution $P_{in}(k)$. Once every gene has been provided with a set of regulators, the dynamics of the network are given by the simultaneous updating of all the gene states according to 
\begin{equation}
 \sigma_i(t+1) = F_i \left(\sigma_{i_1}(t),\sigma_{i_2}(t),\dots,\sigma_{i_{k_i}}(t)\right),
\label{eq:dynamics}
\end{equation}
where $F_i$ is a regulatory function, specific to the gene $\sigma_i$, that is constructed according to the activatory and inhibitory nature of the regulators of $\sigma_i$. 

The differences in the implementation of the Boolean approach mentioned before are related to the way in which the regulatory functions $F_i$ are constructed.\footnote{There is  another important difference in the Boolean implementation which is not related to the regulatory functions but that is worth mentioning, which is the synchronous versus the asynchronous updating schemes. Throughout this work we will use synchronous updating because we want to focus on the differences regarding the construction of the regulatory functions.} For instance, the regulatory functions used in \cite{8,9} for the \emph{A. thaliana} flower development network were carefully constructed taking into account the current biological knowledge about the \emph{combinatorial} action of the regulators on their target genes. This combinatorial action takes place, for instance, with dual regulators whose inhibitory or activatory nature on the target gene depends upon the presence or absence of other regulators (which may compete for the same binding site in the promoter region) \cite{combinatorial}. In contrasts, the regulatory functions used in \cite{11} and \cite{12} for the cell cycle networks of \emph{S. cerevisiae} and \emph{S. pombe} are threshold functions similar to the ones used in artifical neural networks \cite{21,22}. These two schemes, combinatorial functions vs. threshold functions, are very different not only mathematically, but in their very nature. For the use of threshold functions requires the strong assumption that \emph{the effect of activatory and inhibitory regulations, rather than combinatorial, is simply additive}.  

In spite of this strong assumption, Boolean models with threshold functions seem to predict the correct biological sequence of events in the cell cycles of \emph{S. cerevisiae} and \emph{S. pombe} \cite{11,12}. The dynamics in each of these systems exhibit one big attractor that corresponds to the experimentally observed stable state at the end of the cell cycle. This result suggests that, under certain conditions, gene regulatory interactions can indeed be considered as purely additive. In such cases, Boolean models with threshold functions are useful to describe real genetic networks and understand their dynamical properties \cite{13}. Therefore, a thorough study of these kind of mathematical models is necessary. However, although Boolean networks with threshold functions have been extensively studied in the context of spin glasses \cite{15}-\cite{14} and artificial neural networks \cite{21,22}, their dynamical properties in the context of gene regulation are largely unknown. For only the most simple cases of fixed connectivities, equal activator/repressor sterngths and proportions, and fixed threshold values have been explored \cite{13,20}.\footnote{Usually, for spin glasses and neural networks the nodes $\sigma_i$ take the values $\{+1,-1\}$ (rather than $\{1,0\}$). Although models using the spin-like values $\{+1,-1\}$ can be mapped onto models using $\{1,0\}$, the mapping requires a fine tuning of the threshold values $\theta_i$.} 

In this work we investigate the generic dynamical properties of Boolean networks with threshold functions. Our main goal is to compare the behavior of these threshold networks with the one that is already known for standard random Boolean networks (also termed Kauffman networks), focusing on the properties that are relevant to gene regulation processes. To this end, we use different connectivities, activator/repressor strengths and proportions, and threshold values.  In Sec. \ref{sec:2} we describe the Boolean threshold network model and present examples of strong deviations from the ``normal'' behavior observed in Kauffman Boolean networks. Next, in Sec. \ref{sec:3} we use the annealed approximation \cite{28} and the average influence \cite{30} of these networks to calculate the phase diagram for the different parameters involved. In Sec. \ref{sec:4} we present numerical evidence to support the analytical results and discuss the case where anomalies between the theoretical prediction and the numerical simulations arise. Finally, we discuss and summarize our results, highlighting their implications in terms of the applicability of threshold networks for the modeling of gene regulation.

\section{The Boolean threshold network model} \label{sec:2}

\subsection{Definition and general properties} \label{sec:2.1}

In what follows we will refer to Boolean networks with threshold functions as \emph{Boolean threshold networks} (or BTN's). Since threshold functions are a subset of the general class $\mathcal B$ comprising all possible Boolean functions, it is clear that BTN's are a subset of the ensemble of random Boolean networks (RBN's) introduced by Kauffman, in which the regulatory functions $F_i$ are randomly chosen from $\mathcal B$. In the context of gene regulation, the dynamics of BTN's are given by 
\begin{equation} \label{eq:2.1}
  \sigma_i (t+1) = F_i\left(\sigma_{i_1}(t),\ldots,\sigma_i(t),\dots,\sigma_{i_{k_i}}(t)\right) = 
  \begin{cases} 
  1, \quad & \displaystyle{\sum_{j=1}^{k_i}} a_{i,j}\sigma_{i_j}(t) > \theta_i\\
  0, \quad & \displaystyle{\sum_{j=1}^{k_i}} a_{i.j}\sigma_{i_j}(t) < \theta_i\\
  \sigma_i(t), \quad & \displaystyle{\sum_{j=1}^{k_i}} a_{i,j}\sigma_{i_j}(t) = \theta_i,
  \end{cases}
\end{equation}
where $\left\{\sigma_{i_1}, \ldots, \sigma_{i_{k_i}}\right\}$ are the $k_i$ regulators of $\sigma_i$. The interaction strength (or weight) $a_{i,j}$ takes a positive  (or negative) value if $\sigma_{i_j}$ is an activator (or a repressor) of $\sigma_i$, respectively.\footnote{Ofcourse, $a_{i,j}=0$ if there is no interaction between $\sigma_{i_j}$ and $\sigma_i$.} The activation threshold $\theta_i$  of  $\sigma_i$ indicates the minimum value of the sum required for the activation of the node to take place. The dynamic rule given in Eq.~\eqref{eq:2.1} is the same as the one used in Refs. \cite{11,12}. However, in that work the authors considered the simple case in which $a_{i,j}=1$ for activators, $a_{i,j}=-1$ for repressors, and $\theta_i=0$ for almost all the nodes except by a few ones. Additionally, ``self-degradation" was introduced to some of the nodes just by making $a_{i,i}=-1$.

The number $k_i$ of regulators for each node $\sigma_i$ is drawn from a probability distribution $P_{in}(k)$, and then these regulators $\left\{\sigma_{i_1}, \ldots, \sigma_{i_{k_i}}\right\}$ are randomly chosen from anywhere in the system. Each regulatory interaction strength $a_{i,j}$ is set activatory with probability $p$ and inhibitory with probability $1-p$. All activatory interactions have a value $a_{i,j} = a_G$, whereas the inhibitory interactions have a value $a_{i,j}=-a_R$, where $a_G$ and $a_R$ are positive integers. The ratio $a_G/a_R$ measures the relative importance of activation over repression. Thus, if $a_G/a_R$ is small, then inhibitory interactions are dominant, whereas if $a_G/a_R$ is large, then activation dominates over repression. Finally, we set a fixed value of the activation threshold $\theta_i = \theta$ for all nodes. We consider three cases corresponding to three different threshold values: $\theta = 0.5$, $\theta = 0$ and $\theta = -0.5$. The rationale for this choice is two fold. First, these values suffice to illustrate the effects of integer and non-integer thresholds on the dynamics. And second, because these are the values that have been used in models of real genetic networks, obtaining good agreement with experimental observations \cite{11,12}. 

On thing that should be noted from Eq.~\eqref{eq:2.1} is the effect that the value of the threshold $\theta_i$ has on the dynamics. If we consider only integer values for the interaction strenghts $a_{i,j}$, then the equality in Eq.~\eqref{eq:2.1} can be attained only if $\theta_i$ is also an integer. In such a case,  the last row on the right-hand side of Eq.~\eqref{eq:2.1} implies that every node regulates itself. In other words, given the interaction strengths $a_{i,j}$ and the thresholds $\theta_i$, the right-hand side of Eq.~\eqref{eq:2.1} can be written as a Boolean function $F_i$ only if we assume that $\sigma_i$ belongs to its own set of regulators. Because of this, we have explicitely written $\sigma_i(t)$ as one of the arguments of the regulatory function $F_i$. This self-regulation does not necessarily happen in Kauffman Boolean networks, and it can make a big difference with regard to the dynamical behavior. As we will see below, the fact that integer threshold values allow the node $\sigma_i$ to simply stay in their previous state and essentially freeze plays a mayor part in the dynamical behaviour of the network and in its use for biological modeling.

Note also from Eq.~\eqref{eq:2.1} that all the information necessary for the network dynamics is contained in a $N$-dimensional vector $\vec{\theta}=(\theta_1,\theta_2,\dots,\theta_N)$ whose components are the thresholds, and a $N\times N$ matrix $\mathbf{A}$. This matrix is such that $[\mathbf{A}]_{i,j}=a_{i,j}$ if $\sigma_{i_j}$ is a regulator of $\sigma_i$, and $[\mathbf{A}]_{i,j}=0$ otherwise. This is very different from what happens in RBN's, where to store all the information necessary for the network dynamics we need a $N\times N$ matrix containing the topology of the network, and a Boolean function $F_i$ for each node $\sigma_i$. Each of these functions has $2^{k_i}$ entrances, one for each configuration of its $k_i$ inputs. As mentioned before, for a given set of thresholds and interaction strengths, we can also create a Boolean function corresponding to the rule given in Eq.~\eqref{eq:2.1}. However, this limits the set of possible Boolean functions that can be obtained.

\subsection{The Derrida map : Deviations from the Kauffman behaviour} \label{sec:Derrida_map}

One of the most useful ways to study the general dynamics of Boolean networks has been in terms of the propagation of perturbations (also called damage spreading) throughout the network. To this end, let us denote as $\Sigma_t$ the dynamical configuration of the network at time $t$, that is, $\Sigma_t=\left\{\sigma_1(t), \sigma_2(t), \ldots, \sigma_N(t) \right\}$. Let $\Sigma_0$ and $\widetilde{\Sigma}_0$ be two slightly different intitial configurations, namely, $\widetilde{\Sigma}_0$ is almost identical to $\Sigma_0$ except by a few nodes which have reversed their values (this is the initial perturbation or the initial damage). Under the dynamics given in Eq.~\eqref{eq:2.1}, each of these initial configurations will generate a trajectory throughout time:
\begin{eqnarray*}
 &&\Sigma_0\rightarrow\Sigma_1\rightarrow\cdots\rightarrow\Sigma_t\rightarrow\cdots\\
 &&\widetilde{\Sigma}_0\rightarrow\widetilde{\Sigma}_1\rightarrow\cdots\rightarrow\widetilde{\Sigma}_t\rightarrow\cdots
\end{eqnarray*}
These two trajectories may eventually converge (the initial perturbation disappears), diverge (the initial perturbation amplifies), or remain  ``parallel'' (the initial perturbation neither grows nor disappears). These three different behaviors determine the dynamical regime in which the network operates: In the ordered regime the two trajectories typically converge after a transient time. In the chaotic regime the system becomes very sensitive to small changes in the initial condition and the two trajectories diverge from each other. The intermediate case where, on average, perturbations retain their same size corresponds to the so called critical regime. 
The critical regime has been extensively studied and appears to be characteristic property of genetic networks \cite{31,32,33,34,35}.

We quantify the propagation of perturbations in the network in terms of the time evolution of normalized Hamming distance $h(t)$, which is defined as
\begin{equation} \label{eq:3.1}
  h(t) = d \left(\Sigma_t, \widetilde{\Sigma}_t \right)=\frac{1}{N}\sum_{i=1}^{N} |\sigma_i(t)-\widetilde{\sigma}_i(t)|.
\end{equation} 
The assymptotic value $\displaystyle h_\infty = \lim_{t\to\infty}h(t)$ is the final size of the avalanche of perturbations and acts as the order parameter of the system: In the ordered regime $h_{\infty}=0$, while in the chaotic regime $h_{\infty}>0$. In the critical regime $\displaystyle \lim_{t\to\infty}h(t) = 0$ only marginally, which means that it can take a long time for a small perturbation to disappear. 

For a given network realization, $h_\infty$ can be computed numerically in two different ways. The first way is a direct implementation of the definition. We start out the dynamics from two different initial conditions $\Sigma_0$ and $\widetilde{\Sigma}_0$, and let the system evolve for a long time $t_r$. Then, $h_\infty$ is the Hamming distance $h(t_r)$ between the two final configurations $\Sigma_{t_r}$ and $\widetilde{\Sigma}_{t_r}$, averaged over many pairs of initial conditions. We will denote the value of the order parameter obtained by this method as $h_\infty^{(1)}$.

The second way to compute $h_\infty$ is by means of the so-called Derrida map $M\left(h\right)$ \cite{39}, which relates the size of a perturbation avalanche between two consecutive time steps, that is, $h(t+1)=M \left(h(t)\right)$. Starting from two different initial configurations whose Hamming distance is $h_0$, succesive iterations of this map eventually converge to $h_\infty$. Thus, $h_\infty$ is the stable fixed point of the Derrida map: $h_{\infty}=M \left(h_{\infty}\right)$. For RBN's, mean-field theory computations show that $M\left(h\right)$ is a continuous convex monotically increasing function
with the properties $M(0)=0$ and $M(1)<1$. For threshold networks this mapping is still continuous and satisfies $M(0)=0$ and $M(1)<1$, but  it is not clear whether or not it is a monotonically increasing function. Nonetheless, for the set of parameters we use in this work $M\left(h\right)$ seems to satisfy all the properties predicted by the mean-field theory. The fulfillment of these properties is important because this guarantees the existence of one and only one stable fixed point. In this case, the dynamical regime in which the network operates is determined by the slope at the origin of  $M\left(h\right)$, called the \emph{average network sensitivity} $S$: 
\begin{equation}\label{eq:3.3}
 S=\left.\frac{dM(h)}{dh} \right|_{h=0}.
\end{equation}
If $S<1$ then $h_\infty=0$ and the system is in the ordered phase, whereas if $S>1$ then $h_\infty > 0$ and the system is in the chaotic regime. The critical regime is attained for $S = 1$, which is the point at which the phase transition between the ordered and chaotic regimes occur \cite{39,29}. 

To compute $M\left(h\right)$ numerically  for a given network realization, we start from two different configurations $\Sigma_0$ and $\widetilde{\Sigma}_0$ separated by a Hamming distance $h_0$. Next, we evolve these two initial configurations just one time step and compute the Hamming distance $h_1$ between the resulting configurations $\Sigma_1$ and $\widetilde{\Sigma}_1$. The value $M\left[h_0\right]$ of the Derrida map at $h_0$ is then obtained by averaging $h_1$ over many pairs of initial conditions whose Hamming distance is $h_0$. By doing this for all values of $h_0\in(0,1)$ we can construct the full curve $M(h)$ and compute its fixed point $h_\infty$. We will denote the value of the order parameter obtained by this method as $h_\infty^{(2)}$.

For general RBN's it has been shown that $h_{\infty}^{(1)}$ and $h_{\infty}^{(2)}$ are very close to each other. Actually, in the thermodynamic limit $N \rightarrow \infty $ they are the same \cite{39}. The reason for this is that in RBN's the temporal correlations between two consecutive configurations $\Sigma_t$ and $\Sigma_{t+1}$ are inversely proportional to the number of nodes $N$. Therefore, for large networks with completely random Boolean functions the mean-field conditions are satisfied and the temporal evolution is essentially dependent on the previous time step only. However, when temporal correlations extend over several time steps, the Derrida map does not accurately predict the value of the order parameter. In such cases  $h_{\infty}^{(1)}$ and $h_{\infty}^{(2)}$ can differ by a large ammount. This non-ergodic behavior in the network dynamics has been observed in Boolean networks in which only a small subset of the class of all Boolean functions are used \cite{37}, such as canalyzing functions \cite{30} and threshold functions with equal values and proportions of activation/repression strengths \cite{20}.   For the general case of RTN's we also observe a large deviation from the ergodic behavior assumed by the mean-field computation.

\begin{figure}[t]
  \begin{center}
\includegraphics[width=1\textwidth]{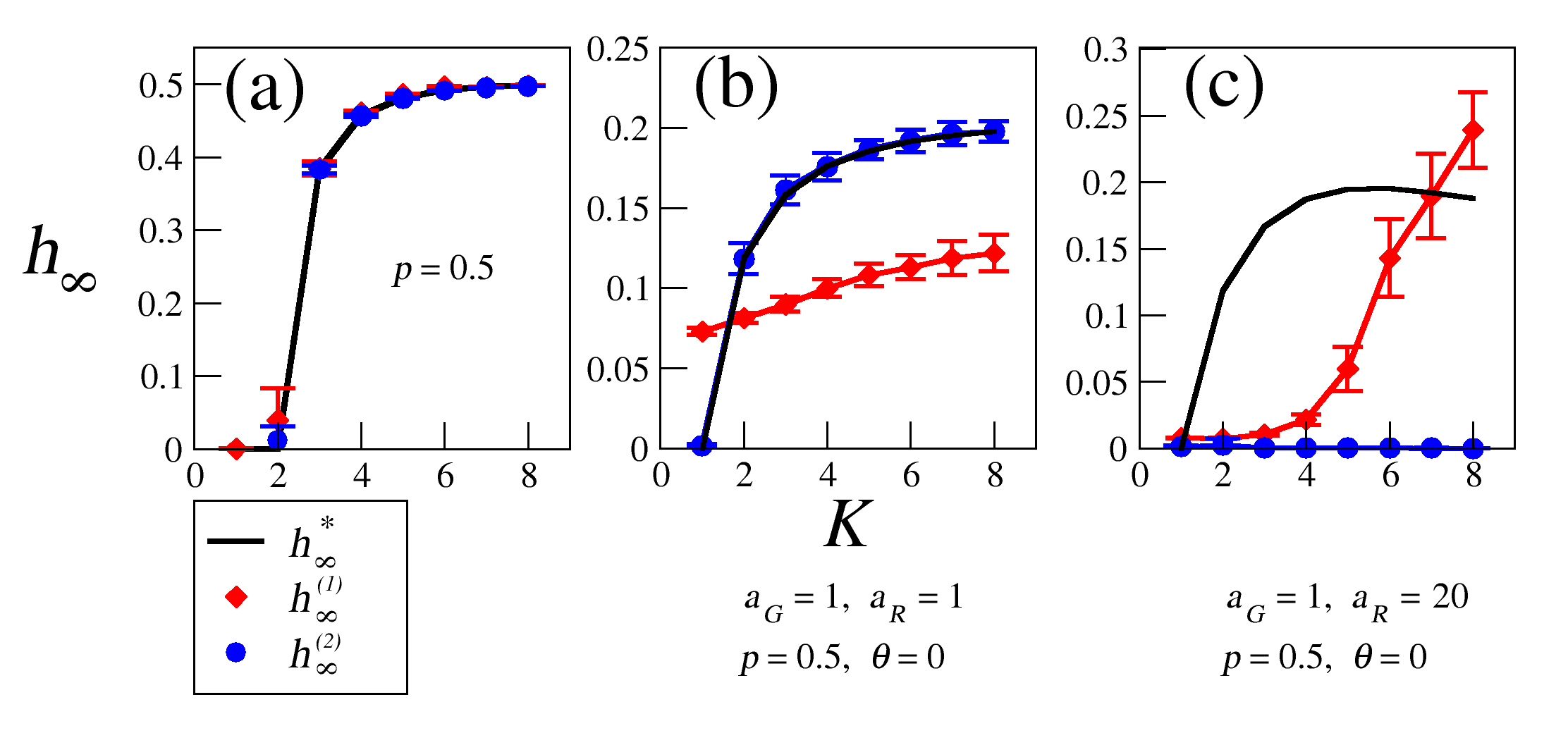}
  \end{center}
  \caption{The nonergodicity of the system is illustrated by plotting the order parameter $h_\infty^{(1)}$ computed  directly from the definition (red diamonds), and the order parameter $h_\infty^{(2)}$ computed as the fixed point of the Derrida map (blue circles). The analytic prediction $h_\infty^*$ from the annealed computation presented in see Sec.~\ref{sec:3} is also plotted (solid line). Three different ensembles of networks are used: (a) Standard Kauffman networks (RBN's). In this case  $h_\infty^{(1)}=h_\infty^{(2)}=h_\infty^*$, which shows that RBN's are ergodic. (b) Random Thresdhold Networks (RTN's) with $p=0.5$, $a_G=a_R=1$ and $\theta=0$. Note in this case that $h_\infty^{(1)}\neq h_\infty^{(2)}$ (although $h_\infty^{(2)}=h_\infty^*$), which reflects the nonergodicity of the dynamics. Finally, (c) corresponds to RTN's with $p=0.5$, $a_G=1$, $a_R=20$ and $\theta=0$. In this last case, not only is $h_\infty^{(1)}\neq h_\infty^{(2)}$, but also the analytic prediction $h_\infty^*$ fails completely. In all cases, each point is the average over 100 network realizations, each having $N=1000$ nodes. For each of these networks we used 10000  pairs of random initial conditions.}
  \label{fig:deviations}
\end{figure}

 In Fig.~\ref{fig:deviations} we plot $h_{\infty}^{(1)}$ (diamonds) and $h_{\infty}^{(2)}$ (circles) as functions of the network connectivity $K$, for RBN's (Fig.~\ref{fig:deviations}a) and RTN's (Fig.~\ref{fig:deviations}a,c). We also plot the quantity $h_{\infty}^*$ predicted analytically using the annealed approximation presented in Sec.~\ref{sec:3}, which is a generalization of the one reported in Ref.~\cite{20}.\footnote{This computation incorporates in an approximate way the temporal correlations between succesive network states using the final number of active and inactive nodes.} Note from Fig.~\ref{fig:deviations}a that for RBN's the three values of $h_{\infty}$ are identical within numerical accuracy, which reflects the ergodicity of the system in this case. However, for RTN's such ergodicity dissappears, as it is apparent from the fact that $h_\infty^{(1)}$ is quite different from $h_\infty^{(2)}$. Fig.~\ref{fig:deviations}b corresponds to the case in which $\theta=0$ for all nodes and the weights take the values $a_{i,j}= \pm 1$, chosen with equal probability. This strong deviation is surprising, especially since it happens for the simplest case similarly to the one used in Refs.~\cite{11,12} for the modelling of the yeast cell-cycle networks. Furthermore, departure from ergodicity is even worse for unequal weights, as shown in Fig. \ref{fig:deviations}c, where the negative weights were chosen to be ten times stronger than the positive weights, i.e. $a_R=10$ and $a_G=1$. In this case $h_\infty^{(1)}$ does not only  deviates from $h_\infty^{(2)}$, but also the analytical prediction $h_\infty^*$ completely fails to reproduce $h_\infty^{(1)}$. The above results indicate that some care must be taken when choosing the parameters in RTN's if these networks are to be used for biological modelling. We explore this issue furtherly in the next sections. 

\section{The phase diagram} \label{sec:3}

The Derrida map $M(h)$ can be computed analytically within the context of the so-called \emph{annealed approximation}, first introduced by Derrida and Pomeau \cite{28}. This mean-field technique assumes statistical independence between the nodes and neglects the temporal correlations developed throughout time between succesive states of the network. The annealed approximation has been successfully used in RBN's to obtain analytically where the phase transition occurs for different topologies and network parameters \cite{29,7}. However, the mean-field assumptions fail dramatically for RTN's as it is illustrated in Fig.~\ref{fig:deviations}. In an attempt to improve the annealed approximation, one has to incorporate into the analysis the temporal correlations between succesive network states \cite{30,37}. This has been done for particular values of the parameters \cite{20}. Here we present a generalization of this computation valid for different activator/repressor strengths, ratios and thresholds.

We start the computation of $M(h)$ by introducing the quantity $I^{(k_d)}$, known as the \emph{influence of $k_d$ variables}. Let us consider an arbitrary network in the ensemble of RTN's, and pick out a node $\sigma_i$ with $k_i$ inputs. Let $\Sigma_t$ and $\widetilde{\Sigma}_t$ be two network configurations in which $k_d$ of the inputs of $\sigma_i$ (with $k_d\leq k_i$) have been damaged, namely, these $k_d$ inputs have opposite values in these two configurations.\footnote{Given that statistical equivalence is assumed, then $\sigma_i$ will be representative of the entire network. Therefore, only the state of the inputs of $\sigma_i$ is important, regardless of the states of all the other nodes.} $I^{(k_d)}$ is defined as the probability that this initial damage of $k_d$ inputs propagates one time step, which means that $\sigma_i$ will have different values in $\Sigma_{t+1}$ and $\widetilde{\Sigma}_{t+1}$. These influences do not only depend on the ensemble of Boolean functions used, but have also been shown to depend heavily on the bias in the expected probability with which the system visits the different states of its configuration space. In previous work, this bias has been expressed in terms of the fraction $b(t)$ of active nodes in the system \cite{20,36,37}. 

By using the annealed approximation assumptions, in \hyperref[Appendix A]{Appendix A} we show that the temporal evolution of $b(t)$ is given by
\begin{subequations}
\label{eqs:mapping_b}
\begin{equation} \label{eq:3.7a}
b(t+1)=B\left(b(t)\right)=p_+\left(b(t)\right) + b(t) \cdot p_0\left(b(t)\right),
\end{equation}
where $p_+$ and $p_0$ are the probabilities that, for a given node, the sum of its inputs is larger than or equal to the threshold $\theta$, respectively. These probabilities can be written as (see \hyperref[Appendix B]{Appendix B})
\begin{align} \label{eq:3.7b}
  p_+\left(b(t)\right)&=\sum_{k_i=1}^{\infty} P_{in}(k_i) \sum_{i=0}^{k_i} \binom{k_i}{i}(1-b)^i b^{k_i-i} \sum_{l=l_i}^{k_i-i} \binom{k_i-i}{l} p^l q^{k_i-i-l},\\
& \quad \mbox{where}\ \ l_i = \left[\frac{(k_i-i)a_G+\theta}{(a_G+a_R)}\right]+1 \nonumber
\end{align}
\begin{align} \label{eq:3.7c}
  p_0\left(b(t)\right)&=\sum_{k_i=1}^{\infty} P_{in}(k_i) \sum_{i=0}^{k_i} \binom{k_i}{i}(1-b)^i b^{k_i-i} \sum_{l=0}^{k_i-i} \binom{k_i-i}{l} p^l q^{k_i-i-l} \nonumber \\& \times \delta_{a_Gl,a_R(k_i-i-l)+\theta}.
  \end{align}
\end{subequations}
Since we are interested in the asymptotic value $h_\infty$ of the order parameter, it is necessary to compute the final number of active elements $b_\infty=\lim_{t\to\infty}b(t)$. This is just the stable fixed point of the map given in Eq. \eqref{eq:3.7a}, namely, $b_{\infty}=B\left(b_{\infty}\right)$. From the set of Eqs.~\eqref{eqs:mapping_b}, the value of $b_\infty$ is computed numerically for each particular realization of parameters. Once the value of $b_\infty$ is known, it is used to compute the influence $I^{(k_d)}$. In \hyperref[Appendix C]{Appendix C} we show that $I^{(k_d)}$ is then given by 
\begin{align} \label{eq:3.4}
  I^{(k_d)}&=  \sum_{i=0}^{k_i} \binom{k_i}{i}p^i q^{k_i-i} \sum_{l=0}^{i} \sum_{m=0}^{k_i-i} \binom{i}{l} \binom{k_i-i}{m} b_{\infty}^{l+m} \left(1-b_{\infty}\right)^{k_i-l-m} \nonumber \\ &\times\mathcal{I}(k_i,k_d,i,l,m),
\end{align}
where $\mathcal{I}(k_i,k_d,i,l,m)$ is defined as
\begin{align} \label{eq:3.6}
  \mathcal{I} &= \sum_{u=u_0}^{u_f} \sum_{v=v_0}^{v_f} \sum_{w=w_0}^{w_f} \frac{\binom{l}{u}\binom{m}{v}\binom{i-l}{w}\binom{k_i-i-m}{k_d-u-v-w}}{\binom{k_i}{k_d}} 
   \nonumber \\ & \times\left\{ H\left(a_G(l-u+w)-a_R(m-v+z)-\theta\right) \cdot \left[(1-b_{\infty})\delta_{a_Gl-a_Rm,\theta} \right. \right. \nonumber \\
 &  \left. + H(a_Rm+\theta-a_Gl)\right]+
  \delta_{a_G(l-u+w),a_R(m-v+z)+\theta} \cdot \left[h(t)\delta_{a_Gl-a_Rm,\theta}\right. \nonumber \\
&+ b_{\infty}H(a_Rm+\theta-a_Gl)  
  \left. +(1-b_{\infty})H(a_Gl-a_Rm-\theta)\right] \nonumber \\ &+ 
  H\left(a_R(m-v+z)+\theta-a_G(l-u+w)\right) \cdot \left[b_{\infty}\delta_{a_Gl-a_Rm,\theta} \right. \nonumber \\
  &\left. \left.  + H(a_Gl-a_Rm-\theta)\right] \right\}.
\end{align}
where the summation is done between the limits of a multivariate hypergeometric distribution, \footnote{Specifically we have that $u_0=\max(0,k_d+l-k_i)$, $u_f=\min(l,k_d)$; $v_0=\max(0,k_d-u-(k_i-l-m))$, $v_f=\min(l,k_d-u)$; $w_0=\max(0,k_d-u-v-(k_i-l-m-i+l))$, $w_f=\min(l,k_d-u-v)$} and $H(x)$ is the Heaviside step function with $H(0)=0$. 

Note that the influence $I^{(k_d)}$ already contains information about the temporal correlations through the value of $b_\infty$. However, the above expressions are not exact because $b_\infty$ is computed from Eqs.~\ref{eqs:mapping_b}, which were formulated using the mean-field assumptions. In spite of this approximation, it is an improvement over the original annealed approximation which completely neglects the temporal correlations.  Once the value of $I^{(k_d)}$ if obtained from the above equations, it is used to obtain the Derrida map, which determines the temporal evolution of the Hamming distance, as \cite{30}
\begin{equation}\label{eq:3.2}
  h(t+1)=M\left(h(t)\right)=\sum_{k_i=1}^{\infty} P_{in}(k_i) \sum_{k_d=0}^{k_i} I^{(k_d)} \binom{k_i}{k_d} \left[h(t)\right]^{k_d} \left[1-h(t)\right]^{k_i-k_d}.
\end{equation}
This equation tells us that the size of a perturbation avalanche after one time step depends on the probability to find $k_d$ damaged input nodes between two configurations $\Sigma_{t}$ and $\widetilde{\Sigma}_{t}$, and on the probability $I^{(k_d)}$ that this damage spreads to the configurations $\Sigma_{t+1}$ and $\widetilde{\Sigma}_{t+1}$ at the next time step. 

\subsection{Sensitivity and influence of 0 variables} \label{sec:3.1}

Once $b_{\infty}$ is obtained from Eq.~\eqref{eq:3.7a}, it is possible to calculate the phase diagrams for the different parameters involved in a network realization using Eqs.~\eqref{eq:3.4}, \eqref{eq:3.2} and \eqref{eq:3.3}. There is however one last point that needs to be considered before the computation of the sensitivity, which is that the average influence of $0$ variables $I^{(0)}$ is not necessarily null.

By definition, $I^{(0)}$ is the probability that, for a given node $\sigma_n$, a damage on none of its input elements spreads to the next time step. This means that $\sigma_n$ will have different values in the configurations $\Sigma_{t+1}$ and $\widetilde{\Sigma}_{t+1}$ even when all of its inputs had the same values in the previous configurations $\Sigma_{t}$ and $\widetilde{\Sigma}_{t}$. In Kauffman RBN's this cannot happen, because the equality of the inputs of $\sigma_n$ in the two configurations $\Sigma_{t}$ and $\widetilde{\Sigma}_{t}$ guarantees that $\sigma_n$ will have the same value in the next configurations $\Sigma_{t+1}$ and $\widetilde{\Sigma}_{t+1}$, and therefore, in this case $I^{(0)}=0$.\footnote{This is why in Ref.~\cite{30} the summation over $k_d$ excludes $k_d=0$, whereas in our Eq.~\eqref{eq:3.2} the sum starts from $k_d = 0$.}. However, for RTN's, the last line in Eq.~\eqref{eq:2.1} makes it possible for $\sigma_n$ to be different in $\Sigma_{t+1}$ and $\widetilde{\Sigma}_{t+1}$ even when all of its inputs were the same in the previous configurations $\Sigma_{t}$ and $\widetilde{\Sigma}_{t}$. This happens when $\sum_{j}a_{n,j}\sigma_{n_j}(t)=\theta$ and $\sigma_n$ had different values in the configurations $\Sigma_t$ and $\widetilde{\Sigma}_t$. In such a case $\sigma_n$ will remain different in the next configurations $\Sigma_{t+1}$ and $\widetilde{\Sigma}_{t+1}$. This can be considered as a damage spread for zero input variables, and consequently $I^{(0)}\neq0$. Note that this happens only when the equality in the last line in Eq.~\eqref{eq:2.1} is satisfied, which in turn occurs only for integer values ot $\theta$.

Taking the above considerations into account, it is possible to write $I^{(0)}$ as 
\begin{equation} \label{eq:3.8}
  I^{(0)}=p_0\left(b_{\infty}\right)\,\ h(t) \quad \hbox{with }\theta \in {\mathbb Z},
\end{equation}
Using the previous equation, one is finally able to get the sensitivity of the network, defined in Eq.~\eqref{eq:3.3}, as
\begin{equation} \label{eq:3.10}
  S= p_0\left(b_{\infty}\right) + \sum_{k_i=1}^{\infty} P_{in}(k_i)\,k_i\,I^{(1)}.
\end{equation}
The derivation of the last two expressions is presented in  \hyperref[Appendix D]{Appendix D}.

Note that $I^{(1)}$ also depends on $k_i$, and that both $I^{(1)}$ and $p_0\left(b_{\infty}\right)$ depend on the parameters of the network realization $p$, $a_G$, $a_R$ and $\theta$. Interestingly, the sensitivity $S$, and thus, the dynamical phase in which the system operates, only depends on the lower influences $I^{(0)}$ and $I^{(1)}$. This means that the effect of small changes in the two configurations $\Sigma_{t}$ and $\widetilde{\Sigma}_{t}$ are the ones that determine the newtork dynamical regime. However, $\Sigma_{t}$ and $\widetilde{\Sigma}_{t}$ cannot be arbitrary, since they must have a fraction of active elements close to the final one, $b_{\infty}$. This restriction has profound effects on the initial apparent dynamical behavior of the network as compared to what actually happens at the end of the dynamics, in the sense that two trajectories that initially appear to converge may end up diverging, and vice versa. We will  discuss this problem in Sec.~\ref{sec:4}.

\subsection{The phase diagram for the homogeneous random topology} \label{sec:3.2}

Eq.~\eqref{eq:3.10} determines the structure of the phase diagram as a function of the network topology (contained in $P_{in}(k)$), and the other parameters of the system. Here we consider the homogeneous random topology $P_{in}(k) = \delta_{K,k}$ in which each node has exactly $K$ regulators randomly chosen from anywhere in the system. In such a case, Eq.~\eqref{eq:3.10} establishes a relationship between the sensitivity $S$ and the value of the parameters $K$, $p$,  $a_G$, $a_R$ and $\theta$. The ordered phase occurs in those regions of the parameter space in which $S<1$, whereas the chaotic phase occurs whenever $S>1$. The critical region is the one for which $S=1$. As the parameter space is 5-dimensional, an exhaustive exploration is neither illustrative nor computationally feasible. Instead, we present the phase diagram $K$ vs. $p$  for the following cases, which are representative of the general behavior observed across the entire parameter space: 
\begin{itemize}
 \item {\bf Case 1:} Activating and inhibiting interactions are of the same magnitude ($a_G=1$, $a_R=1$);
 \item {\bf Case 2:} Inhibiting interactions are stronger than activating ones ($a_G=1$, $a_R=2$);
 \item {\bf Case 3:} Activating interactions are stronger than inhibiting ones ($a_G=2$, $a_R=1$);
 \item {\bf Case 4:} Inhibiting interactions are always dominant ($a_G=1$, $a_R=20$);
 \item {\bf Case 5:} Activating interactions are always dominant ($a_G=20$, $a_R=1$);
\end{itemize}
Additionally, for each of the five cases listed above we used the threshold values $\theta = 0.5$, $\theta = 0$ and $\theta = -0.5$. 

\begin{figure}
  \begin{center}
\includegraphics[width=1\textwidth]{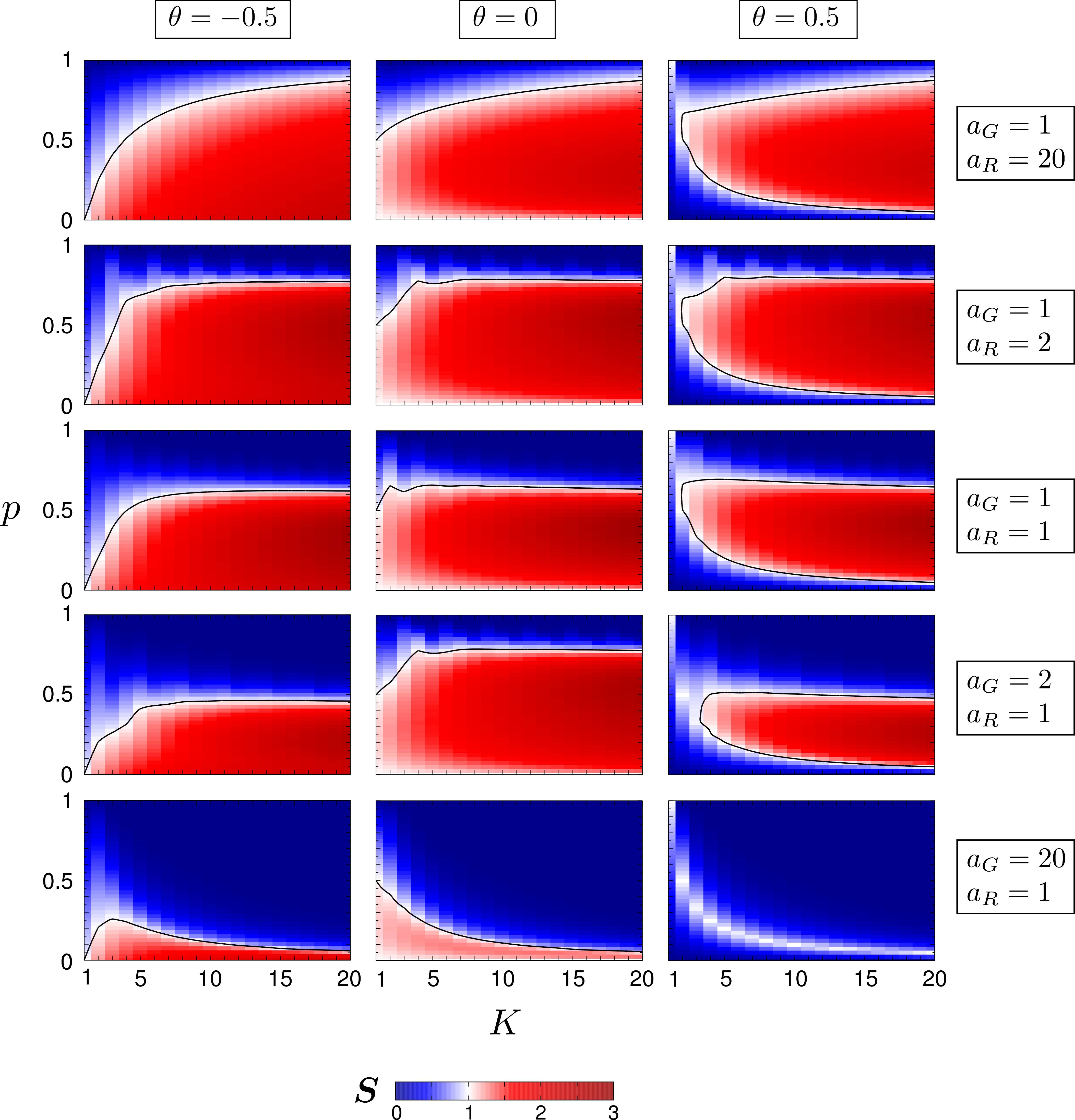}
  \end{center}
  \caption{Phase diagram $p$ vs. $K$ for threshold networks with different parameters, obtained by numerically solivng Eq. \eqref{eq:3.10}. The color code represents the  value of the sensitivity $S$. The ordered phase ($S<1$) is represented in blue while the chaotic phase ($S>1$) is represented in red. Zones near the critical region ($S=1$) are white, while the critical region itself is represented by the black line. Note the asymmetry of the phase diagrams, especially for $\theta=0$ and $\theta=-0.5$.}
  \label{fig:phase_diagrams}
\end{figure}

The resulting phase diagrams are shown in Fig.~\ref{fig:phase_diagrams}. It is immediately apparent from this figure the asymmetric structure of the phase diagram with respect to the activator fraction (measured by $p$) and strength (measure by the quotient $a_G/a_R$). In general, it appears that activators strongly push the network into the frozen phase (in blue), while repressors move it towards the chaotic phase (in red), but less drastically. This can be seen in the extreme cases of dominant activators ($a_G=20$, $a_R=1$) where the chaotic region almost dissappears, while for the opposite case of dominant repression ($a_G=1$, $a_R=-20$) the frozen region is considerably smaller than the chaotic one. Another important point to note is the different behavior of the critical line for the three threshold values of interes: For $\theta=0.5$ there are two critical values of $p$ for each value of $K$, whereas for $\theta=-0.5$ and $\theta = 0$ there is only one. In this sense, of all the cases shown in Fig.~\ref{fig:phase_diagrams}, the phase diagrams for $\theta=0.5$ are the ones closer to the phase diagram obtained for RBN's \cite{7}.

Finally, it is important to mention that we obtain the same results reported in Ref.~\cite{20} for the special case $p=0.5$, $a_G=1$ and $a_R=1$, but only for the threshold values $\theta=0.5$ and $\theta=-0.5$. However, for $\theta=0$ we obtain a completely different behavior as the one reported in  Ref.~\cite{20}. Indeed, we find that the phase transition occurs at $K=1$, whereas the authors in Ref.~\cite{20} report that the phase transition ocurs between $K=12$ and $K=13$. This discrepancy is due to not properly taking into account the self-regulation conveyed in the last line of Eq.~\eqref{eq:2.1}, which happens only for integer threshold values  (see  Sec.~\ref{sec:2.1}). In the next section we present numerical results that support our analytic approach for integer threshold values.

\section{Numerical experiments} \label{sec:4}

To test the validity of the expressions obtained in Sec.~\ref{sec:3} we performed numerical simulations of the network dynamics using ensembles of 100 RTN's, with $N=1000$ nodes each. We used networks with homogeneous random topologies and varied $K$ from $K=1$ to $K=8$. The other parameters  $p$, $a_G$, $a_R$ and $\theta$ were chosen to represent the different behaviors depicted in the phase diagrams shown in Fig.~\ref{fig:phase_diagrams}. 

\subsection{The uncorrelated sensitivity $S_0$ and the final avalanche size $h_{\infty}$} \label{sec:4.1}

To compare the analytical expressions with the results of the numerical simulation we need to compute two parameters: The \emph{uncorrelated network sensitivity} $S_0$ and the \emph{final avalanche size} $h_{\infty}^{(1)}$. In Sec.~\ref{sec:Derrida_map} we describe how to compute $h_\infty^{(1)}$ for a given network realization. To compute $S_0$ let us consider two initial configurations $\Sigma_0$ and $\widetilde{\Sigma}_0$ that differ only in one element, namely, whose Hamming distance is $1/N$:
\begin{equation}
 d\left(\Sigma_0,\widetilde{\Sigma}_0\right)=\frac{1}{N}.
\label{eq:hamming_1}
\end{equation}
Then, $S_0/N$ is the average Hamming distance of the two configurations at the next time step:
\[
 S_0 = N \left\langle d\left(\Sigma_1,\widetilde{\Sigma}_1\right)\right\rangle,
\]
where the average $\langle \cdot \rangle$ is taken over all possible pairs of initial conditions satisfying Eq.~\eqref{eq:hamming_1}. In other words, $S_0$ is the average size of the perturbation avalanche after one time step, given an initial perturbation of only one node. Note that $S_0$ is the slope at the origin of the Derrida map without taking into account the correlations developed throughout time between network states. This is the reason why we call $S_0$ the ``uncorrelated'' sensitivity. Therefore, for large $N$, $S_0$ should be the sensitivity of the network given by Eq.~\eqref{eq:3.10} with $b_{\infty}=b_0=0.5$, which assumes complete independence between the network states.\footnote{Since the initial configuration $\Sigma_0$ in Eq.~\eqref{eq:hamming_1} is chosen randomly from all possible configurations, the sequence of 0's and 1's in $\Sigma_0$ can be thought of as $N$ independent Bernoulli trials with probability $1/2$, which gives  $b_0=0.5$ for the  expected fraction of $1$'s.}

\begin{figure}[t]
  \begin{center}
\includegraphics[width=0.7\textwidth]{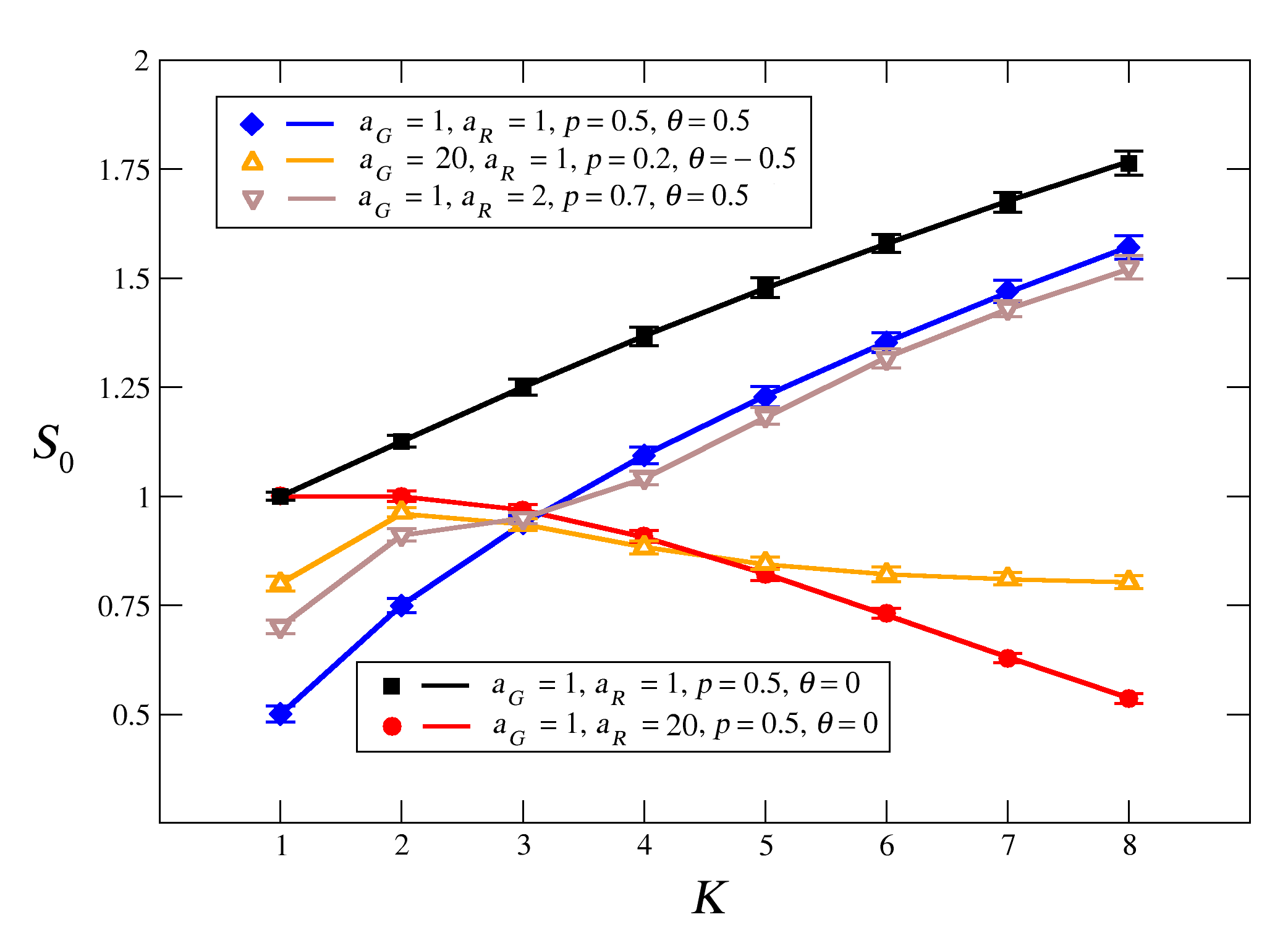}
  \end{center}
  \caption{Uncorrelated sensitivity $S_0$ for RTN's with $N=1000$ and different values of the parameters $p$, $a_G$, $a_R$ and $\theta$. The symbols are numerical data  computed using ensembles of 100 networks. For each of these networks, $S_0$ was averaged over $10000$ pairs of initial conditions differing in just one node. The error bars represent the standard deviation. The solid lines correspond to the theoretical result gien in Eq.~\eqref{eq:3.10} with $b=0.5$. Note the excellent agreement between the theoretical prediction and the numerical data for all the different parameters used.}
  \label{fig:sensitivities}
\end{figure}

\begin{figure}[t]
  \begin{center}
\includegraphics[width=1\textwidth]{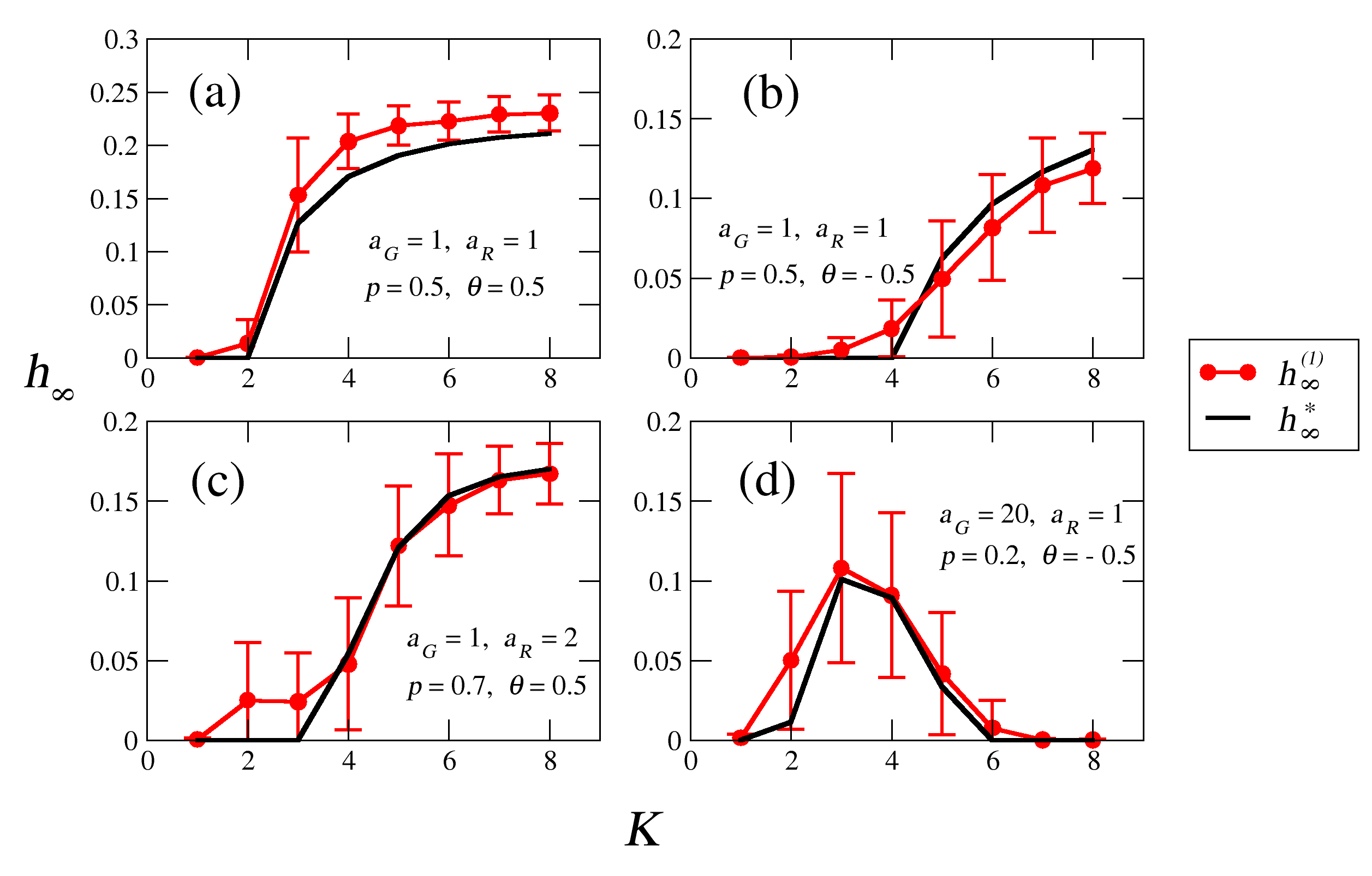}
  \end{center}
  \caption{Final size of the perturbation avalanche computed numerically ($h_\infty^{(1)}$, red circles), and analytically as the fixed point of Eq.~\eqref{eq:3.2} ($h_{\infty}^*$, solid black line). The numerical data were computed for an ensemble of $100$ RTN's  with $N=1000$ and different values of the parameters $p$, $a_G$, $a_R$ and $\theta$. The important point in this figure is the use of noninteger threshold values: $\theta=\pm0.5$. For each network realization, we averaged $h_\infty^{(1)}$ over $10000$ pairs of random initial conditions with Hamming distance $h_0=0.1$. Note the large standard deviations in the numerical dada (error bars). Despite this enormous variability in each network realization, the average numerical data qualitatively follow well the theoretical prediction.}
  \label{fig:h_infty-vs-K}
\end{figure}

Since the uncorrelated sensitivity $S_0$ has no dependence on the correlations developed in time, we expect the analytic results derived in Sec.~\ref{sec:3} to accurately reproduce the behavior of $S_0$. However, we do not expect this analytic computation to describe as acurately the value of $h_{\infty}^{(1)}$, because in this computation the temporal correlations were approximately taken into account only through the value of $b_\infty$, whereas the actual value of $h_{\infty}^{(1)}$ depends on the precise way in which the network evolves in time. This is illustrated in Figs.~\ref{fig:sensitivities} and \ref{fig:h_infty-vs-K}.  

It can be seen in Fig.~\ref{fig:sensitivities} that the uncorrelated sensitivity $S_0$ computed numerically (symbols) shows a remarkable agreement with the theoretical prediction (solid line) for the different combinations of parameters used. It is interesting to note the variety of behaviours exhibitted by $S_0$ in RTN's, which is in marked constrast with the linear behavior observed in standard Kauffman Nets. Indeed, for RBN's $S_0=2p(1-p)K$, whereas for Kauffman networks with canalyzing functions $S_0=1/2+(K-1)/4$ \cite{30}. In contrast, Fig.~\ref{fig:sensitivities} shows that for RTN's the dependance of $S_0$ on $K$ is nonlinear and can even change inflexion or decrease with increasing $K$. This general nonlinear behaviour occurs even for the simple cases $p=0.5$, $a_G=a_R=1$, $\theta=0$, and $\theta=\pm0.5$, where we have (see \hyperref[Appendix E]{Appendix E} for a derivation)
\begin{equation}
 S_0 = \left\{
	      \begin{array}{lcl}
	       \displaystyle{\frac{K+1}{2^{2K}}\binom{2K}{K}} & \mbox{for} & \theta = 0 \vspace{0.1in} \\ 
	       \displaystyle{\frac{K}{2^{2K}}\binom{2K}{K}} & \mbox{for} & \theta = \pm0.5  \\ 
	      \end{array}
      \right.
\label{eq:4.3}
\end{equation}
in which $S_0\sim \sqrt{K}$ for large $K$. The above result allows the network to ramain close to the critical phase for a wider range of values of $K$ than standard RBN's. This might be important given that there is evidence showing that real genetic networks, in which the gene input connectivity varies considerably from one gene to another, work near the critical phase $S_0=1$ \cite{35}.

With regard to the final size of the perturbation avalanche,  Fig.~\ref{fig:h_infty-vs-K} shows the value $h_\infty^{(1)}$ computed numerically (line with symbols), and the value $h_\infty^{*}$ predicted by the analytic computation of Sec.~\ref{sec:3} (solid line).  Although $h_\infty^{(1)}$ and $h_\infty^{*}$ are qualitatively very similar to each other for the cases $\theta=\pm0.5$ depicted in Fig.~\ref{fig:h_infty-vs-K}, their quantitative correspondence is not as precise as it was for $S_0$. As it was mentioned before, this lack of precision was expected due to the approximation in the computation of the temporal correlations. However, for integer threshold values  $h_\infty^{(1)}$ and $h_\infty^{*}$ do not necessarily agree even qualitatively, as it is illustrated in Fig.~\ref{fig:deviations} for the special case $\theta=0$. We discuss the origing of this discrepancy further below. In the mean time, it is important to emphasize the reason why the Derrida map is not always useful to discriminate the dynamical regime in which the network operates. 

Fig.~\ref{fig:h_t-vs-t} shows the temporal evolution of the average Hamming distance $h(t)$ between two trajectories that started from two slightly different initial conditions $\Sigma_0$ and $\widetilde{\Sigma}_0$. (The average is taken over many pairs of initial conditions.) In all the cases shown in this figure, $h(t)$ decreases in the first time steps. Therefore, according to the Derrida map, which takes into account only the first time step, these networks should be in the ordered regime. However, after that initial decrease, the Hamming distance $h(t)$ increases again reaching a value considerably larger than the initial Hamming distance $h(0)$. Thus, in the long term the initial perturbation is amplified, which means that the dynamical regime in which these networks operate turns out to be chaotic. It should be noted that the behavior reported in Fig.~\ref{fig:h_t-vs-t} was obtained for networks with biologically reasonable values of the parameters: $a_G=a_R=1$, $\theta=0.5$, $p=0.5$ and $K=3$. 

\begin{figure}
  \begin{center}
\includegraphics[width=0.6\textwidth]{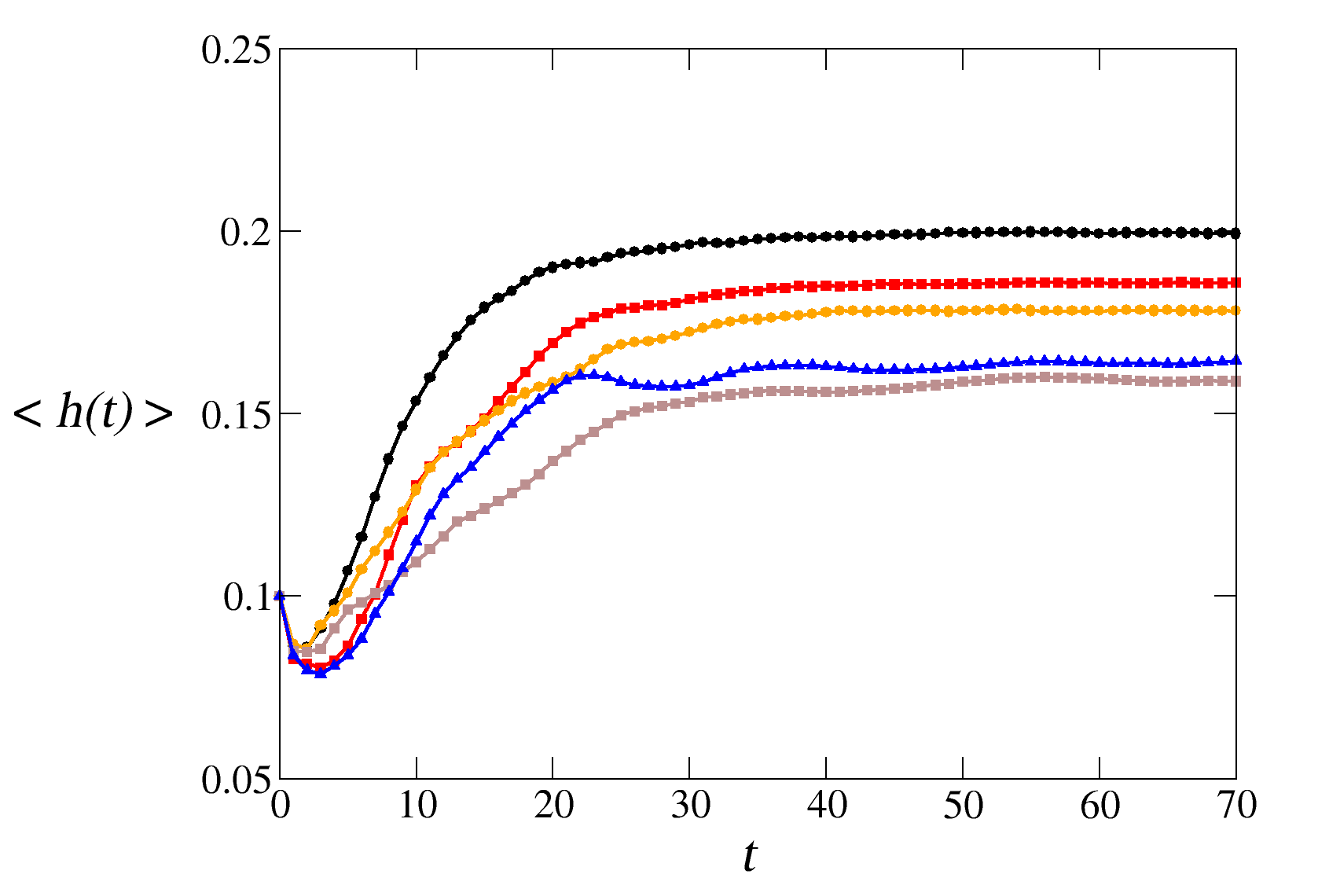}
  \end{center}
  \caption{Temporal evolution of the Hamming distance for 5 different random threshold network realizations with $N=1000$, $a_G=a_R=1$, $\theta=0.5$, $p=0.5$ and $K=3$. Each curve is the average over $10000$ randomly chosen pairs of initial conditions separated by a Hamming distance $h_0=0.1$. Note that initially the Hamming distance decreases. Using only the Derrida map, which takes into account only the first time step, one would conclude that the networks operate in the ordered regime. However, the correlations developed in time due to the network structure and the number of active nodes make the Hamming distance rise again and approach a nonzero value, which is characteristic of the chaotic regime.}
\end{figure}

\subsection{The $\theta=0$ case} \label{sec:4.2}

\begin{figure}
  \begin{center}
\includegraphics[width=0.6\textwidth]{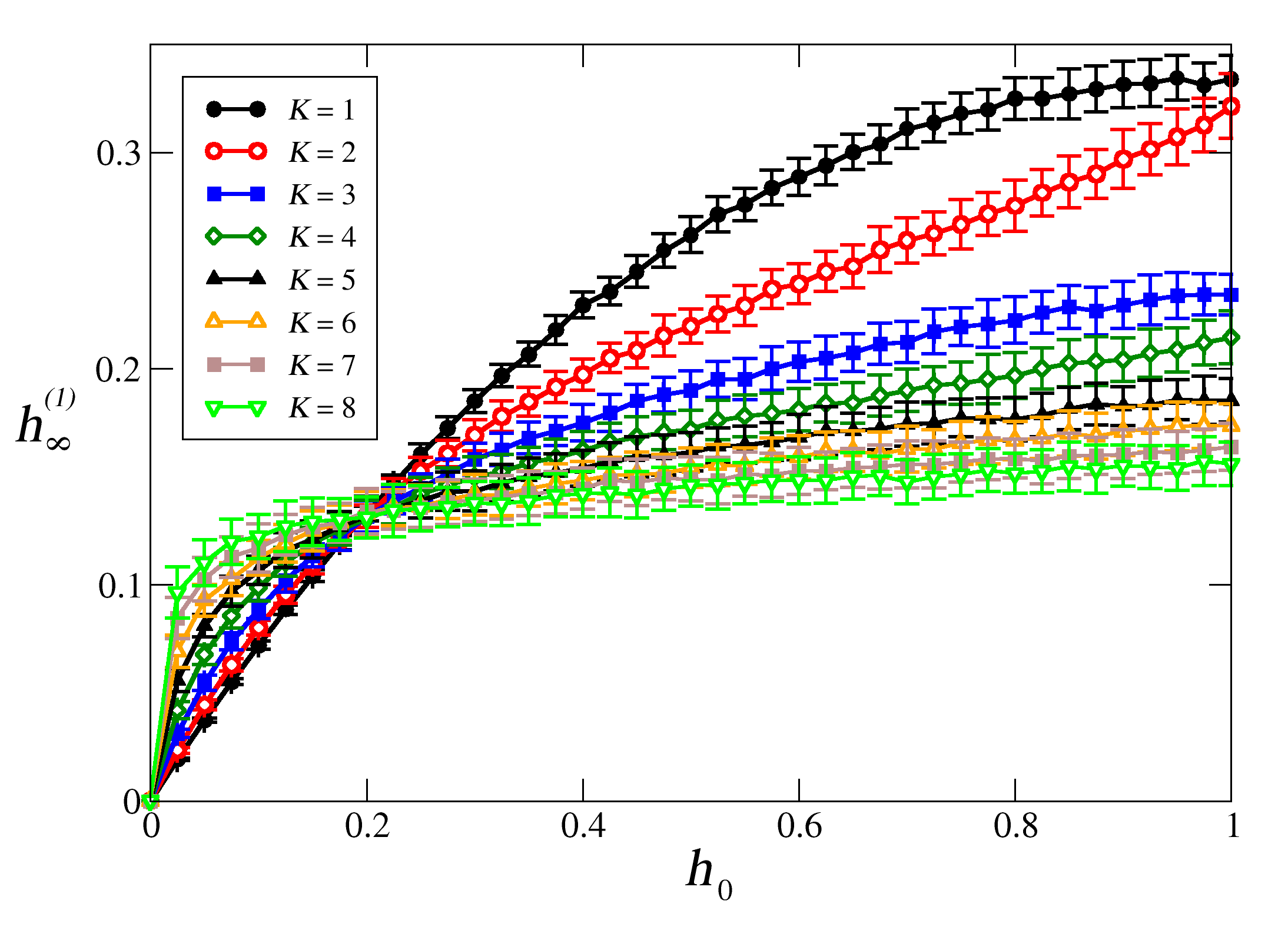}
  \end{center}
  \caption{Final size $h_\infty^{(1)}$ of the perturbation avalanche as a function of the initial perturbation size $h_0$. Each point is the average over $100$ random threshold network realizations with $N=1000$, $a_G=a_R=1$, $p=0.5$, $\theta=0$, and $1000$ pairs of random initial conditions for every $h_0$ in each of these networks. Note the strong dependence of $h_\infty^{(1)}$ on $h_0$, especially for small values of $K$ where the temporal correlations are stronger. For large values of $K$ these correlations become weaker and consequently $h_\infty^{(1)}$ becomes almost independent of $h_0$.}
  \label{fig:h_t-vs-t}
\end{figure}

We now address the anomalous case $\theta=0$. As we have seen in the previous section, in this case the annealed approximation gives very accurate results for the initial sensitivity $S_0$ but very poor results for the final avalanche size $h_{\infty}^{(1)}$. As discussed in Sec.~\ref{sec:2}, integer thresholds allow the possibility for some nodes to become frozen whenever their input sum in Eq.~\eqref{eq:2.1} equals the threshold. These frozen nodes generate explicit correlations in time, which in turn produce a strong dependance on the history of the dynamics, and thus, on the initial conditions. This is illustrated in Fig.~\ref{fig:h_t-vs-t} where the final avalance size $h_\infty^{(1)}$ is plotted against the initial perturbation size $h_0=h(0)$ for networks with $p=0.5$, $a_G=a_R=1$ and $\theta=0$. It is apparent from this figure that for $K\leq5$ the value of $h_\infty^{(1)}$ strongly depends on $h_0$, and this dependece becomes less strong as $K$ increases. This is because for large values of $K$ it is harder for the input sum to equal the threshold.

\begin{figure}
  \begin{center}
\includegraphics[width=1\textwidth]{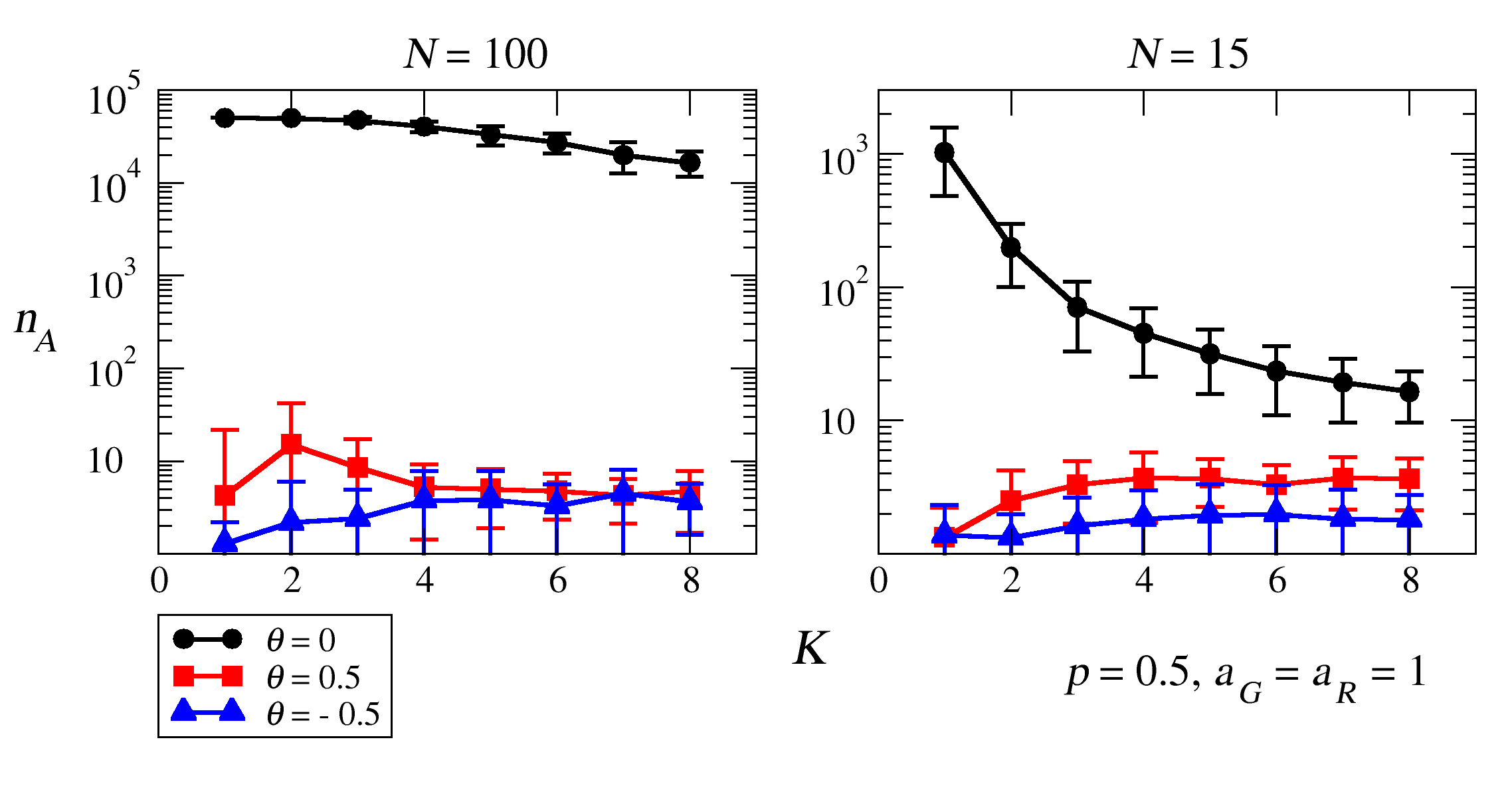}  
\end{center}
  \caption{Average number of attractors as a function of the network connectivity $K$ for RTN's with (a) $N=100$  and (b) $N=15$ nodes. In each panel, we used $p=0.5$, $a_G=a_R=1$ and $\theta=0$ (circles), $\theta=0.5$ (squares) and $\theta=-0.5$ (triangles). For the large networks in (a) we sample the configuration space using  $50,000$ randomly chosen initial conditions, whereas in (b) the full configuration space was probed. Both in (a) and (b) each point is the average over ensembles of 50 networks. Note the extremely large number of attractors obtained for $\theta=0$, especially for moderately small values of $K$. In particular,  for $\theta=0$ in (a) almost every sampled initial condition leads to a different attractor.}
  \label{fig:attractors} 
\end{figure}   

\begin{figure}
  \begin{center}
\includegraphics[width=1\textwidth]{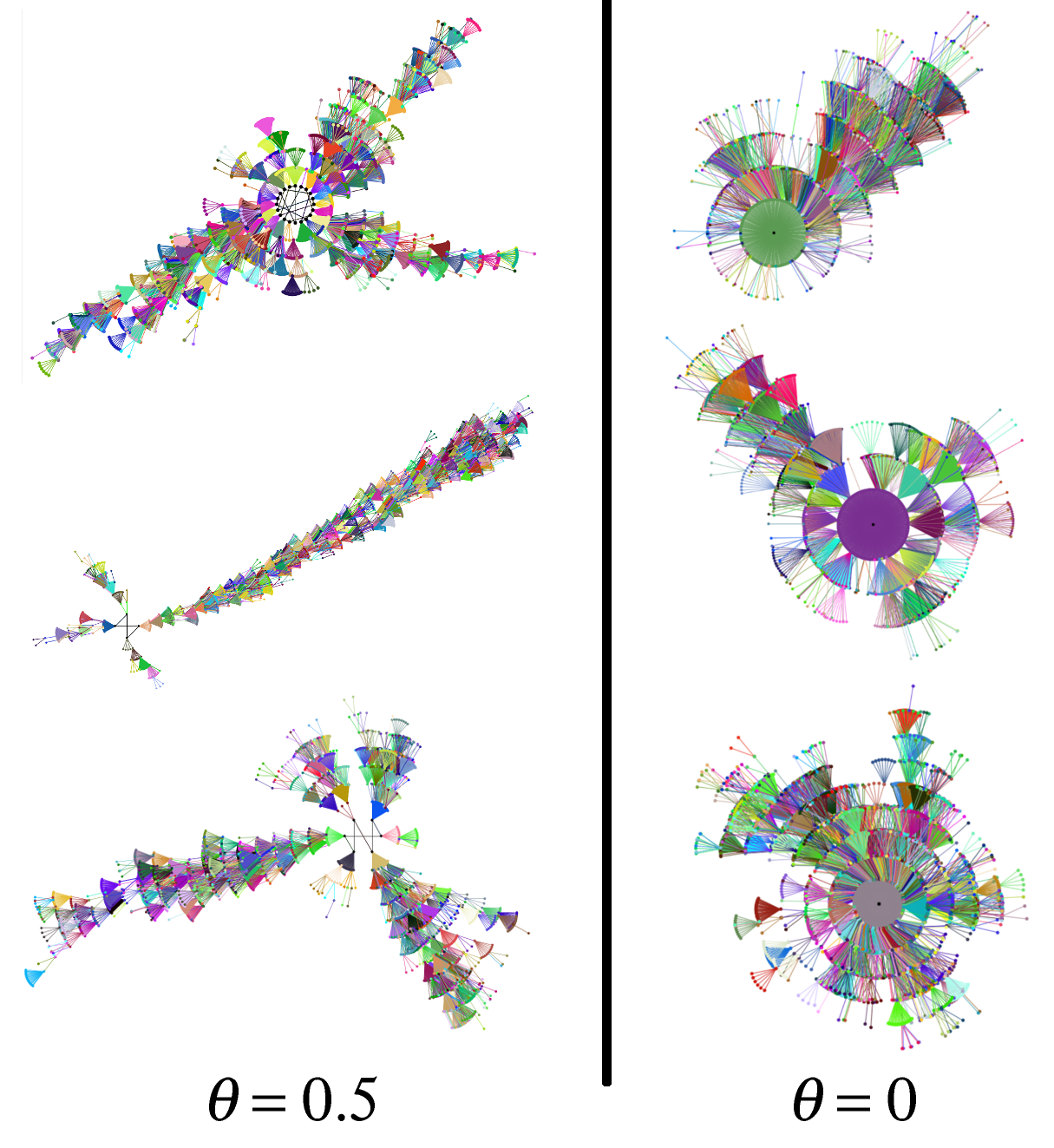}  
\end{center}
  \caption{Structure of the attractor landscape of RTN's with $N=15$, $K=8$, $p=0.5$, $a_R=a_G=1$, and $\theta = 0.5$ (left) and $\theta=0$ (right). Each of the basins of attraction shown is the largest one in a given network realization. Note that for $\theta=0.5$ the attractors  have several configurations (black dots at the center of each structure), whereas for $\theta = 0$ all the atractors are punctual (only one dot at the center). Additionally, for $\theta=0.5$ the long arm-like structures indicate the existence of long transients and a reduced number of configurations which all the routes that go to the attractor have to go through. Contrary to this, for $\theta=0$ the basins of attraction look more sparse and with shorter transients and more distributed across the configuration space. }
  \label{fig:basins}
\end{figure}    

Another problematic consequence of using integer threshold values is the existence of an enormous number of punctual attractors, many of which differ only in the value of just one node. This anomaly has been noted before in Ref.~\cite{20}. It also occurs in the cell cycle models of \emph{S. cerevisiae} \cite{11} and \emph{S. pombe} \cite{12}, where many punctual attractors with small basins of attraction were found.\footnote{However, in these cell cycle models the authors deemed the attractors with small basins of attractions as biologically irrelevant and neglected them.}  Fig.~\ref{fig:attractors} shows the average number of attractors as a function of the network connectivity $K$ for $\theta = 0$ (circles), $\theta=0.5$ (squares) and $\theta = -0.5$ (triangles). In all cases we used $p=0.5$ and $a_G=a_R=1$. Fig.~\ref{fig:attractors}a corresponds to large networks with $N=1000$. In this case the number of possible configurations is astronomically huge ($\Omega = 2^{1000}$). Therefore, an under sampling of the state space has to be done, in which case only some attractores will be found. We sampled $5\times10^4$ configurations. Surprisingly, for $\theta=0$ \emph{almost every sampled initial configuration ended up in a different attractor}. The same happens for smaller networks with $N=100$, as it is shown in Fig.~\ref{fig:attractors}b. However, the same undersampling performed in networks with non-integer threshold values ($\theta=\pm0.5$) reveals a number of atractors which is several orders of magnitude smaller than the one obtained for $\theta=0$. Finally, Fig.~\ref{fig:attractors}c shows similar results but for small networks with $N=15$, for which the entire state space can be probed ($\Omega=2^{15}=32768$). Note that in this case, the average number of attractors decreases with $K$ for $\theta=0$. This behavior is marked contrast with the one observed for non-integer threshold values (and for RBN's), where the average number of attractors grows with the network connectivity $K$.

The $\theta=0$ case presents ``anomalous'' behavior not only with regard to the number of attractors, but also in the structure of the state space.\footnote{Here, by ``anomalous'' we mean with respect to what is observed in standard Kauffman networks.}   Fig.~\ref{fig:basins} shows the largest basin of attraction for three network realizations with $N=15$, $K=8$, $p=0.5$, $a_R=a_G=1$, and $\theta=0.5$ (left part) and $\theta = 0$ (right part). For those parameters the networks are in the chaotic regime. It can be seen from this figure that the structure of the largest basin of attraction for the non-integer threshold is characterized by long transients and attractor length. This is similar to what is obtained using RBN's with the same $p$ and the same $K$. However, for $\theta=0$ the structure is quite different. Note first that all the attractors are punctual. Although this is not the rule, it is the most probable situation for $\theta=0$. Additionally, the transients are comparatively short and the whole basins look somehow sparse as compared to the ones on the left part. The long arm-like structures observed in the basins of attraction for $\theta=0.5$ reflect that the routes to reach the attractor are concentrated in a few number of states. From a biological point of view, these few states can be considered as the ``checkpoints'' of the differentiation (or metabolic) pathway. Contrary to this, the sparseness observed in the basins of attraction for $\theta=0$ indicate that the routes to reach the attractor are much more distributed across the state space, and therefore, the existence of ``checkpoints'' is harder to attain.         

\section{Summary and discussion} \label{sec:5}

We have investigated the dynamical properties of random threshold networks (RTN's), which differ from standard Kauffman networks (or random Boolean networks, RBN's) in that the regulation of the state of the nodes is done by means of threshold functions. These networks have been used in the modeling of genetic regulatory networks of real organisms using parameter values that seem biologially meaningful \cite{11,12}. An important characteristic of the threshold network model is that it assumes that gene regulation is an additive process. Namely, that the combined effect of the regulators of a given gene on the state of that gene is just the sum of the positive regulations minus the negative ones. Because of its simplicity, this assumption is very tempting when constructing models of gene regulatory networks. However, there are many examples in real systems showing that gene regulation is combinatorial rather than additive, which means that the effect of some regulators (i.e. whether activatory or inhibitory) depends on the presence or absence of some other regulators.\footnote{An common example of this are dual regulators in \emph{E. coli} \cite{38}.} In several cases, these combinatorial processes are represented by Boolean functions that cannot be obtained from threshold functions (for instance, the XOR function), which makes the additivity assumption mentioned above questionable. 

Additionally, a deeper analysis of the dynamics of RTN's reveals anomalies inconsistent with the expected behavior of gene regulation models, precisely for the ``biologically meaningful'' values of the parameters that have been used. Of particular importance is the case of integer threshold values (illustrated here using $\theta=0$), where the networks typically have an enormous amount of attractors. Also in this case, there is a sharp disagreement between the ensemble properties predicted by the annealed approximation and the ones observed in the numerical simulation using concrete network realizations. It is worth emphasizing that this disagreement happens neither for RTN's with non-integer threshold values nor for RBN's. It is not surprising to find such a disagreement in specific networks constructed in very peculiar ways (as the ones used for the modeling of the cell-cycles). What is surprising is that the typical members of the ensemble, constructed in a completely random way, present such anomalies. In fact, the cell-cycle networks in Refs.~\cite{11,12} do exhibit these anomalies, as they have a large number of attractors. However, the authors of that work considered most of these attractors as biologically irrelevant because of their small basins sizes, and neglected them. Nonetheless, from an evolutionary point of view it is not clear whether or not the size of the basin of attraction is relevant, as it is not known whether this parameter is under selective pressure. Actually, in other studies of gene regulatory networks of real organisms, the biologically meaningful attractors of the wild-type organism do not possess the largest basins of attraction, but on the contrary, sometimes they have very tiny basins \cite{9,10}.

We computed analytically the phase diagram that determines in which parameter region the network has chaotic, ordered or critical dynamics. Contraty to what happens for RBN's, the phase diagrams obtained for RTN's are always asymmetric with respect to the fraction of positive regulations $p$. It is only for non-integer positive thresholds (illustrated here for the case $\theta=0.5$), that the phase diagram looks semi-symmetric, similar to the one obtained for RBN's (see Fig.~\ref{fig:phase_diagrams}). This may be important because in such a case there is a bigger freedom to vary $p$ and still remain close to the critical region, especially at low newtork connectivities  as the ones reported for networks of real organisms that operate in the critical regime ($K\sim2$, \cite{35}). Note that the case $\theta=0.5$ is biologically reasonable not only in terms of the phase diagram, but also because it corresponds to a situation in which at least one positive regulator has to be active in order to activate the target gene. Contraty to this, in the case $\theta=-0.5$ all genes get activated when all of its positive become innactive, which is an artifact of the model rather than a biological observed behavior. Furthermore, the phase transition for $\theta=-0.5$ and $p=0.5$ occurs at a network connectivity $K=4$ (seel Fig.~\ref{fig:h_infty-vs-K}), which is large compared to the one observed in real networks. Therefore, this case with negative threshold values seems to be inadequate for the study of the theoretical properties of gene regulatory networks. Consequently, some of the conclusions about the evolution of RTN's with negative thresholds might have to be reinterpreted \cite{drossel_PRE_2010}.        

One important point analyzed in this work was the usefulness of the Derrida map to elucidate the network's dynamical regime. As it is shown in Fig.~\ref{fig:h_t-vs-t}, for RTN's the first steps in the dynamics may indicate that the network is in the ordered regime, while in fact the long-term behavior is chaotic. This occurs when long-term correlations are developed thoughout time, which always happens in RTN's, especially for integer threshold values. For in such a case, the self-regulation implied by the last line in Eq.~\eqref{eq:2.1} induces long-term memory in the system. For RBN's these long-term correlations do not exist, and the Derrida map accurately predicts the network's dynamical regime. This is an important aspect that has not been properly taken into account in current work that aims to characterize the network's dynamical regime in real biological networks. Therefore, a more thorough study is necessary in this direction.  

Finally, it is important to note that in this work we used the same fixed connectivities, interaction strengths and thresholds for all genes. However, more realistic situations would require assigning these quantities differently to the different genes in the network. For instance, for some genes the inhibitory regulators may be dominant, whereas for other genes the activatory regulators would dominate. Also, genes with integer as well as non-integer threshold values can coexist in the same network. Exploration of these possibilities can reveal dynamical behaviors more consistent with biological systems, which in turn would help to discern the model's characteristics relevant for biological modeling.

\phantomsection
\addcontentsline{toc}{section}{Acknowledgements}
\section*{Acknowledgements} \label{Acknowledgements}

We thank H. Larralde for fruitful discussions. J.G.T. Za\~nudo acknowledges CONACyT for a research asistant scholarship. This work was partially supported by PAPIIT-UNAM grants IN112407-3 and IN109210.

\appendix
\phantomsection
\addcontentsline{toc}{section}{Appendix}
\label{Appendix}
\section*{Appendix}

\setcounter{section}{1}
\setcounter{equation}{0}
\phantomsection
\addcontentsline{toc}{subsection}{Appendix A}
\label{Appendix A}
\subsection*{Appendix A: Derivation of the map $b(t+1)=B\left(b(t)\right)$}

We first remember that $b(t)$ represents the fraction of active nodes ($\sigma_i=1$) for a given network configuration at time $t$. Using the annealed approximation, the evolution of $b(t)$ depends only on the previous state and can thus be given by the map $B$, which relates the number of active nodes after two consecutive time steps.

To explicitly obtain $B$ let us consider the following. Since in the annealed approximation we assume statistical independence between the nodes, the fraction of active elements $b(t)$ can actually be considered as the probability that an arbitrary node $\sigma_i$ is active at time $t$. Therefore, $b(t+1)$ corresponds to the probability that a node is active after one time step. From the dynamical equation for the nodes, Eq.~\eqref{eq:2.1}, it is apparent that there are only 2 ways in which this can happen: Either the sum $\sum_{j}a_{n,j}\sigma_{n_j}(t)$ was larger than $\theta$ or it was equal to $\theta$. In this last case we additionally need the node itself to be active at time $t$ so that it is still active at time $t+1$, which happens with probability $b(t)$. If we denote $p_+\left(b(t)\right)$ as the probability that $\sum_{j}a_{n,j}\sigma_{n_j}(t)>\theta$ and $p_0\left(b(t)\right)$ as the probability that $\sum_{j}a_{n,j}\sigma_{n_j}(t)=\theta$, then $B$ must be the sum of the probabilities of these two events:
\begin{equation} \label{eq:Da}
b(t+1)=B\left(b(t)\right)=p_+\left(b(t)\right) + b(t) \cdot p_0\left(b(t)\right),
\end{equation}
This corresponds to Eq.~\eqref{eq:3.7a} in the main text. The full expression for $p_0$ and $p_+$ are derived in the next section, \hyperref[Appendix B]{Appendix B}. 

\setcounter{section}{2}
\setcounter{equation}{0}
\phantomsection
\addcontentsline{toc}{subsection}{Appendix B}
\label{Appendix B}
\subsection*{Appendix B: Derivation of $p_0\left(b\right)$ and $p_+\left(b\right)$}

To derive these expressions we use the mean-field  method from Ref.~\cite{17}. Let us denote $P_{\Sigma}(y)$ as the probability distribution function of the sum $\xi_i=\sum_{j=1}^{k_i}a_{i,j}\sigma_{i_j}$ in Eq.~\eqref{eq:2.1} of a node $\sigma_i$ with $k_i$ inputs. The probability that $\xi_i=0$ and $\xi_i>0$ are then given, respectively, by 
\begin{align} \label{eq:B.1}
 p_+\left(b,k_i\right)=\lim_{\epsilon\rightarrow0}\int_{\theta+\epsilon}^{\infty} P_{\Sigma}(y) \ dy \\
 \label{eq:B.2} p_0\left(b,k_i\right)=\lim_{\epsilon\rightarrow0}\int_{\theta-\epsilon}^{\theta+\epsilon} P_{\Sigma}(y) \ dy. 
\end{align}
The weights $a_{i,j}$ can be considered as random variables which take the value $a_G$ with probability $p$ and $-a_R$ with probability $q=1-p$, as defined in Sec.\ref{sec:2}. Using this and denoting $b$ as the probability that a node is active, we can consider the $\xi_{ij}=a_{i,j}\sigma_{i_j}$ as random variables which can take the values 0 with probability $1-b$, $a_G$ with probability $bp$ and $-a_R$ with probability $bq$, that is
\begin{equation} \label{eq:B.3}
 P_{\xi}(x)=(1-b)\delta(x)+bp~\delta(x-a_G)+bq~\delta(x+a_R),
\end{equation}
Using the statistical independence assumption of the annealed approximation, each $\xi_{ij}$ in the sum $\xi_i=\sum_{j=1}^{k_i}\xi_{ij}$ is an independent random variable with probability distribution $P_{\xi}(x)$. Because of this, $\xi_i$ is the sum of $k_i$ independent random variables, and thus $P_{\Sigma}(y)$ must be the $k_i$-fold convolution of $P_{\xi}$:
\begin{equation} \label{eq:B.4}
 P_{\Sigma}(y)=\underbrace{P_{\xi}\ast P_{\xi} \ast \cdots \ast P_{\xi} (y)}_\text{$k_i$ times}.
\end{equation}
Taking the Fourier transform of the above equation we get
\begin{equation} \label{eq:B.5}
 \hat P_{\Sigma}(\omega)=\left[\hat P_{\xi} (\omega)\right]^{k_i},
\end{equation}
where $\hat P_{\xi}= (1-b) + bpe^{-i\omega a_G} + bqe^{i\omega a_R}$. Thus, $P_{\Sigma}(y)$ is obtained by taking the inverse transform of the last equation. Using the binomial theorem twice we get
\begin{align}
P_{\Sigma}(y) &= \frac{1}{2\pi}\int_{-\infty}^{\infty} e^{i\omega y}\left[\hat P{\xi} (\omega)\right]^{k_i} \ d\omega \nonumber \\
&= \frac{1}{2\pi} \sum_{i=0}^{k_i} \int_{-\infty}^{\infty} e^{i\omega y} \binom{k_i}{i} (1-b)^i b^{k_i-i} \left(pe^{-i\omega a_G} + qe^{i\omega a_R}\right)^{k_i-i} \ d\omega \nonumber \\
&= \frac{1}{2\pi} \sum_{i=0}^{k_i} \sum_{l=0}^{k_i-i} \binom{k_i}{i} \binom{k_i-i}{l} (1-b)^i b^{k_i-i} p^l q^{k_i-i-l} \\ &\times \int_{-\infty}^{\infty}  e^{i\omega y} e^{-i\omega a_Gl} e^{i\omega a_R(k_i-i-l)} \ d\omega \nonumber \\
&= \sum_{i=0}^{k_i} \sum_{l=0}^{k_i-i} \binom{k_i}{i} \binom{k_i-i}{l} (1-b)^i b^{k_i-i} p^l q^{k_i-i-l} \delta \left[a_Gl-a_R(k_i-i-l) - y\right] \label{eq:B.6}
\end{align}
If we substitute this last result into Eqs. \eqref{eq:B.1} and \eqref{eq:B.2} we find
\begin{align} 
 p_+\left(b,k_i\right)&=\lim_{\epsilon\rightarrow0}\int_{\theta+\epsilon}^{\infty} P_{\Sigma}(y) \ dy \nonumber \\
&= \sum_{i=0}^{k_i} \binom{k_i}{i} (1-b)^i b^{k_i-i} \nonumber \\ & \times \lim_{\epsilon\rightarrow0}\int_{\theta+\epsilon}^{\infty} \left\{ \sum_{l=0}^{k_i-i} \binom{k_i-i}{l} p^l q^{k_i-i-l} \delta \left[a_Gl-a_R(k_i-i-l) - y\right]\right\} \ dy \nonumber \\
&= \sum_{i=0}^{k_i} \binom{k_i}{i}(1-b)^i b^{k_i-i} \sum_{l=l_i}^{k_i-i} \binom{k_i-i}{l} p^l q^{k_i-i-l}, \label{eq:B.7} \\ & \quad \hbox{where } l_i=\frac{(k_i-i)a_G+\theta}{a_G+a_R}+1  \nonumber \\ \nonumber \\
p_0\left(b,k_i\right)&=\lim_{\epsilon\rightarrow0} \nonumber \int_{\theta-\epsilon}^{\theta+\epsilon} P_{\Sigma}(y) \ dy \nonumber \\
&= \sum_{i=0}^{k_i} \binom{k_i}{i} (1-b)^i b^{k_i-i} \nonumber \\ & \times \lim_{\epsilon\rightarrow0}\int_{\theta-\epsilon}^{\theta+\epsilon} \left\{ \sum_{l=0}^{k_i-i} \binom{k_i-i}{l} p^l q^{k_i-i-l} \delta\left[a_Gl-a_R(k_i-i-l) - y\right]\right\} \ dy \nonumber \\
&= \sum_{i=0}^{k_i} \binom{k_i}{i}(1-b)^i b^{k_i-i} \sum_{l=0}^{k_i-i} \binom{k_i-i}{l} p^l q^{k_i-i-l} \delta_{a_Gl,a_R(k_i-i-l)+\theta} \label{eq:B.8}
\end{align}
where in Eq.~\eqref{eq:B.7} the minimum value of $l=l_i$ was chosen so that the argument of the Dirac delta function is always above $\theta$, as specified by the limit. Similarly in Eq.~\eqref{eq:B.7} it is chosen so that it is exactly equal to $\theta$. Finally, since the probability distribution of $k_i$ is given by $P_{in}(k)$ we have that
\begin{align} 
 p_+\left(b,k_i\right)&= \sum_{k_i=1}^{\infty} P_{in}(k_i)~p_+\left(b,k_i\right) \nonumber \\
 &=\sum_{k_i=1}^{\infty} P_{in}(k_i) \sum_{i=0}^{k_i} \binom{k_i}{i}(1-b)^i b^{k_i-i} \sum_{l=l_i}^{k_i-i} \binom{k_i-i}{l} p^l q^{k_i-i-l}, \label{eq:B.9} \\ 
p_0\left(b,k_i\right)&= \sum_{k_i=1}^{\infty} P_{in}(k_i)~p_0\left(b,k_i\right) \nonumber \\
&=\sum_{k_i=1}^{\infty} P_{in}(k_i) \sum_{i=0}^{k_i} \binom{k_i}{i}(1-b)^i b^{k_i-i} \sum_{l=0}^{k_i-i} \binom{k_i-i}{l} p^l q^{k_i-i-l} \nonumber \\ & \times \delta_{a_Gl,a_R(k_i-i-l)+\theta}. \label{eq:B.10}
\end{align}
which correspond, respectively, to Eqs. \eqref{eq:3.7b} and \eqref{eq:3.7c}. 

\setcounter{section}{3}
\setcounter{equation}{0}
\phantomsection
\addcontentsline{toc}{subsection}{Appendix C}
\label{Appendix C}
\subsection*{Appendix C: Derivation of $I^{(k_d)}$ for Boolean threshold networks}

We first remember the definition of $I^{(k_d)}$, the influence of $k_d$ variables. Denoting $\Sigma_t$ and $\widetilde{\Sigma}_t$ as two configurations in which $k_d$ of the inputs of an arbitrary node $\sigma_i$ are different, $I^{(k_d)}$ is the probability that, after a time step, the node $\sigma_i$ in the new configurations, $\Sigma_{t+1}$ and $\widetilde{\Sigma}_{t+1}$, are different from each other. 

Consider the average over all possible active/inactive and activatory/inhibitory configurations of the inputs of an arbitrary node $\sigma_i$. If the node has $k_i$ inputs, activator probability $p$ and a fraction of active nodes $b$, then this average is given by 
\begin{equation} \label{eq:A.0} 
  \langle X \rangle_{IC}=\sum_{i=0}^{k_i} \binom{k_i}{i}p^i q^{k_i-i} \sum_{l=0}^{i} \sum_{m=0}^{k_i-i} \binom{i}{l} \binom{k_i-i}{m} b^{l+m} \left(1-b\right)^{k_i-l-m} X.
\end{equation}
Here $p^i q^{k_i-i}$ ($q=1-p$) is the probability that for a given input configuration with $k_i$ regulators
there are $i$ activatory interactions and $k_i-i$ inhibitory ones, which can be chosen in $\binom{k_i}{i}$
possible ways. $b^{l+m} \left(1-b\right)^{k_i-l-m}$ is the probability that there are $l$
active activatory inputs and $m$ active inhibitory ones, which can be arranged in
$\binom{i}{l} \binom{k_i-i}{m}$ different ways. Since $I^{(k_d)}$ is the probability for one arbitrary node (regardless of the number of active or inactive inputs), we have to compute the average over all possible input configurations of  $\mathcal{I}(k_i,k_d,i,l,m)$, which is the probability that a damage spreads
when $k_d$ of the input elements are damaged given that this configuration has $i$ activatory inputs
and $k_i-i$ inhibitory ones, which in turn have $l$ and $m$ active/inactive input nodes, respectively.

To find $\mathcal{I}(k_i,k_d,i,l,m)$ we need to consider all possible ways in which the
damaged nodes can be arranged. There are $l$ active activatory input nodes and $m$
active inhibitory ones, and thus, $i-l$ inactive activatory elements and $k_i-i-m$
inactive inhibitory ones. Therefore, we may have $u$ damaged active
activatory inputs, $v$ damaged active inhibitory ones, $w$ damage inactive activatory ones and $z=k_d-u-v-w$
damaged inactive ones. Since there are $\binom{l}{u}\binom{m}{v}\binom{i-l}{w}\binom{k_i-i-m}{k_d-u-v-w}$
possible ways in which damage can be distributed, and given that there are $\binom{k_i}{k_d}$ total ways
in which the damaged nodes can be arranged, then the probability for each value of
$u$, $v$, $w$ is given by a multivariate hypergeometric distribution
\begin{equation} \label{eq:A.1}
 \hbox{Pr}(u,v,w)=\frac{\displaystyle{\binom{l}{u}\binom{m}{v}\binom{i-l}{w}\binom{k_i-i-m}{k_d-u-v-w}}}{\displaystyle{\binom{k_i}{k_d}}} \quad \quad \begin{array}{l}
u=u_0,\ldots,u_f \\ v=v_0,\ldots,v_f \\ w=w_0,\ldots,w_f \end{array},
\end{equation}
where
\begin{equation} \label{eq:A.2}
 \begin{array}{l l}
u_0=\max(0,k_d+l-k_i), \quad & u_f=\min(l,k_d) \\
v_0=\max(0,k_d-u-(k_i-l-m)), \quad & v_f=\min(l,k_d-u)\\
w_0=\max(0,k_d-u-v-(k_i-l-m-i+l)), \quad & w_f=\min(l,k_d-u-v).
\end{array}
\end{equation}
Finally, we need to consider all possible ways in which a damage can actually make the state of $\sigma_i$
in the damage and the undamaged configuration different at time $t+1$. From the definitions of $l$ and $m$ it follows that, before damage, there are $l$ active activatory input nodes and $m$ active inhibitory ones. After damage, using the definitions of $u$, $v$, $w$ and $z$, there will be $l-u+w$ active activatory input elements and $m-v+z$ active inhibitory ones. Using this information and Eq.~\eqref{eq:2.1}, it is clear that the damage can spread in 3 different ways:
\begin{subequations}
\begin{description}
  \item[i)] If in the damaged configuration the sum is above the threshold ($a_G(l-u+w)-a_R(m-v+z)>\theta$),
then the state of the nodes can be different if, without the damage the sum is either below the threshold ($a_Gl-a_Rm<\theta$) or exactly at the threshold ($a_Gl-a_Rm=\theta$) but with the condition that it was inactive
at the time-step before (so that they are different in the next time step), which happens with probability $1-b$.
\begin{align} \label{eq:A.3a}
 P_1&=H\left(a_G(l-u+w)-a_R(m-v+z)-\theta\right) \cdot \left[(1-b)\delta_{a_Gl-a_Rm,\theta} \right.\nonumber \\ & \left.+ H(a_Rm+\theta-a_Gl)\right].
\end{align}
  \item[ii)] The second possiblity is that in the damaged configuration the sum is just at the threshold ($a_G(l-u+w)-a_R(m-v+z)=\theta$). In that case damage can spread if: (a) before the damage the sum is above the threshold ($a_Gl-a_Rm>\theta$) but only if it was inactive before (with probability $1-b$); (b) if before the damage the sum is below the threshold ($a_Gl-a_Rm<\theta$) but only if it was active on the step before (with probability $b$); and (c) if before the damage the sum is again exactly at the threshold ($a_Gl-a_Rm=\theta$) but only if both configurations where damaged before (with probability $h(t)$).
\begin{align} \label{eq:A.3b}
 P_{2}&=\delta_{a_G(l-u+w),a_R(m-v+z)+\theta}\left[h(t)\delta_{a_Gl-a_Rm,\theta} + b~H(a_Rm+\theta-a_Gl) \right.\nonumber \\ & \left.+(1-b)H(a_Gl-a_Rm-\theta)\right].
\end{align}
  \item[iii)] In the third possibility the damaged configuration has its sum below the threshold ($a_G(l-u+w)-a_R(m-v+z)<\theta$)
so the state of the nodes will differ if before the damage the sum is either above the threshold ($a_Gl-a_Rm>\theta$)
or exactly at the threshold ($a_Gl-a_Rm=\theta$) but only if it was active
at the time-step before, which happens with probability $b$.
\begin{align} \label{eq:A.3c}
 P_{3}&=H\left(a_R(m-v+z)+\theta-a_G(l-u+w)\right) \cdot \left[b~\delta_{a_Gl-a_Rm,\theta} \right. \nonumber \\ & \left.+ H(a_Gl-a_Rm-\theta)\right].
\end{align}
\end{description}
\end{subequations}
Since all three cases can make damage spread, then the total probability for damage spreading $\mathcal I$
is the sum of these three possiblilities: Eqs. \eqref{eq:A.3a}, \eqref{eq:A.3b} and \eqref{eq:A.3c}, averaged over all possible arrangements of damaged configurations, Eq.~\eqref{eq:A.2},
\begin{align} \label{eq:A.4}
  \mathcal{I} &= \sum_{u=u_0}^{u_f} \sum_{v=v_0}^{v_f} \sum_{w=w_0}^{w_f} Pr(u,v,w)\cdot\left(P_3+P_2+P_3\right) \nonumber \\
   &= \sum_{u=u_0}^{u_f} \sum_{v=v_0}^{v_f} \sum_{w=w_0}^{w_f} \frac{\binom{l}{u}\binom{m}{v}\binom{i-l}{w}\binom{k_i-i-m}{k_d-u-v-w}}{\binom{k_i}{k_d}} 
   \nonumber \\ & \times \left\{ H\left(a_G(l-u+w)-a_R(m-v+z)-\theta\right) \cdot \left[(1-b_{\infty})\delta_{a_Gl-a_Rm,\theta} \right. \right. \nonumber \\
 &  \left. + H(a_Rm+\theta-a_Gl)\right]+
  \delta_{a_G(l-u+w),a_R(m-v+z)+\theta} \cdot \left[h(t)\delta_{a_Gl-a_Rm,\theta}\right. \nonumber \\
&+ b_{\infty}H(a_Rm+\theta-a_Gl)  
  \left. +(1-b_{\infty})H(a_Gl-a_Rm-\theta)\right] \nonumber \\ &+ 
  H\left(a_R(m-v+z)+\theta-a_G(l-u+w)\right) \cdot \left[b_{\infty}\delta_{a_Gl-a_Rm,\theta} \right. \nonumber \\
  &\left. \left.  + H(a_Gl-a_Rm-\theta)\right] \right\}.
\end{align}
Finally, averaging this damage spread for a given input configuration $\mathcal{I}(k_i,k_d,i,l,m)$ over all possible input configurations using Eq.~\eqref{eq:A.0}, we
get the average influence of $k_d$ variables:
\begin{align} \label{eq:A.5}
  I^{(k_d)}&=\left\langle \ \mathcal{I}(k_i,k_d,i,l,m) \ \right\rangle_{IC} \nonumber \\ &= \sum_{i=0}^{k_i} \binom{k_i}{i}p^i q^{k_i-i} \sum_{l=0}^{i} \sum_{m=0}^{k_i-i} \binom{i}{l} \binom{k_i-i}{m} b^{l+m} \left(1-b\right)^{k_i-l-m} \nonumber \\ &\times\mathcal{I}(k_i,k_d,i,l,m).
\end{align}
Eqs. \eqref{eq:A.4} and \eqref{eq:A.5} with $b=b_{\infty}$ correspond to the formulas \eqref{eq:3.6}
and \eqref{eq:3.4}, respectively, in the main text.

\setcounter{section}{4}
\setcounter{equation}{0}
\phantomsection
\addcontentsline{toc}{subsection}{Appendix D}
\label{Appendix D}
\subsection*{Appendix D: Derivation of $S$ and $I^{(0)}$}

As discussed in Sec. \ref{sec:4.2}, $I^{(0)}$ is the probability that, for an arbitrary node $\sigma_i$, damage spreads at the next time step when none of its input elements are different between the two initial configurations, $\Sigma_t,\widetilde{\Sigma}_t$. Because of the possibility of having this sum giving exactly the threshold value $\theta$, $I^{(0)}$ is  zero only for noninteger thresholds.  

The case for integer values of $\theta$ can be obtained from Eqs. \eqref{eq:3.4} and \eqref{eq:3.6}. However, from Eq.~\eqref{eq:2.1} we can see that the only way for a damage to spread when none of the input elements are different is by having $\sum_{j}a_{n,j}\sigma_{n_j}(t)=\theta$. Thus,  $I^{(0)}$
must correspond to the probability $p_0$ that the sum gives exactly the threshold value, Eq.~\eqref{eq:3.7c}. Since we need to consider the nonergodicity of the system, we use the final fraction of activatory nodes $b_{\infty}$ as the value of $b$. Now, for damage to spread not only
does the sum need to be at the threshold, but also the two nodes must be different initially, otherwise they would be the same in the next time-step and the damage would not spread. Since $h(t)$ can be considered as the probability that two arbitrary nodes are different, then $I^{(0)}$ must be the multiplication of both probabilities
\begin{equation} \label{eq:E.1}
  I^{(0)}=p_0\left(b_{\infty}\right)\cdot\ h(t) \quad \hbox{with }\theta \in {\mathbb Z},
\end{equation}
Taking this last result into consideration we can now calculate the sensitivity $S$ of RTN's. Using 
\eqref{eq:3.2} in \eqref{eq:3.3} we get
\begin{equation} \label{eq:E.2}
  S= \sum_{k_i=1}^{\infty} P_{in}(k_i) \left. \left[\frac{dI^{(0)}}{dh}+ k_i(I^{(1)}-I^{(0)}) \right] \right|_{h=0}.
\end{equation}
From Eq.~\eqref{eq:3.6} it is apparent that $I^{(1)}$ does not actually depend on $h$. This happens because the middle term of Eq.~\eqref{eq:3.6}, which is the only one with a $h$ dependance, can never be nonzero for the case of $k_d=1$. This is because $I^{(1)}$ refers to the case where after $k_d$ changes of an arbitrary input, the summation of both configurations is still at the threshold, which cannot happen when only $1$ input is changed. With this information and using Eq.~\eqref{eq:E.1} in \eqref{eq:E.2} we finally find
\begin{align} \label{eq:E.3}
  S &= \sum_{k_i=1}^{\infty} P_{in}(k_i) \left[ \left. p_0\left(b_{\infty}\right) + k_i\left(I^{(1)}-I^{(0)} \right) p_0\left(b_{\infty}\right) \cdot\ h \right] \right|_{h=0} \nonumber \\
 &= \sum_{k_i=1}^{\infty} P_{in}(k_i) \left( p_0\left(b_{\infty}\right)+ k_iI^{(1)} \right) \nonumber \\
&= p_0\left(b_{\infty}\right) + \sum_{k_i=1}^{\infty} P_{in}(k_i) k_iI^{(1)}.
\end{align}
Eqs. \eqref{eq:E.1} and \eqref{eq:E.3} correspond, respectively, to Eqs. \eqref{eq:3.8} and \eqref{eq:3.10} in the main text.

\setcounter{section}{5}
\setcounter{equation}{0}
\phantomsection
\addcontentsline{toc}{subsection}{Appendix E}
\label{Appendix E}
\subsection*{Appendix E: Derivation of $S_0$ for $p=0.5$, $a_G=a_R=1$, $\theta=0$ and  $\theta=\pm0.5$}

We first remember from Sec. \ref{sec:4.1} that $S_0$, the uncorrelated network sensitivity, is the average number of nodes by which two configurations differ after one time step if they initially differed in only one element:
\[
 S_0 = N \left\langle d\left(\Sigma_1,\widetilde{\Sigma}_1\right)\right\rangle, \quad \hbox{with} \quad d\left(\Sigma_0,\widetilde{\Sigma}_0\right)=\frac{1}{N}.
\]
Because of the thermodynamic limit assumed in the annealed approximation, $S_0$ should correspond to the sensitivity of the network given in Eq.~\eqref{eq:3.10} with $b_{\infty}=b_0=0.5$. This last choice of $b$ is a consequence of the initial configurations being chosen randomly. In what follows we consider consider the case ni which $P_{in}(k)=\delta_{k,K}$ with $p=0.5$, $a_G=a_R=1$, $\theta=0$ and $\theta=\pm0.5$.

\phantomsection
\addcontentsline{toc}{subsubsection}{E.1}
\subsubsection*{E.1 \quad $S_0$ for $p=0.5$, $a_G=a_R=1$ and $\theta=0$} \label{sec:E.1}

Using Eqs. \eqref{eq:3.8} and \eqref{eq:3.10} for the integer threshold $\theta=0$, we have
\begin{equation} \label{eq:C.1}
S_0= p_0\left(b_0\right) + KI^{(1)}. 
\end{equation}
where from Eqs. \eqref{eq:3.6} and \eqref{eq:3.7c} 
\begin{align} 
p_0\left(b_0\right)&= \frac{1}{2^{2K}}\sum_{i=0}^{K}\binom{K}{i}2^i \ \sum_{l=0}^{K-i}\binom{K-i}{l} \delta_{l,K-i-l}  \label{eq:C.2},\\ 
I^{(1)}&= \frac{1}{2^{2K}}\sum_{i=0}^{K}\binom{K}{i} \ \sum_{l=0}^{i} \sum_{m=0}^{K-i} \binom{i}{l} \binom{K-i}{m} \mathcal{I}(a_G=1,a_R=1,\theta=0,b=0.5). \label{eq:C.3}
\end{align}
For $\mathcal{I}(a_G=1,a_R=1,\theta=0,b=0.5)$ we consider the following. Since we have an integer threshold $\theta=0$ and we are looking for the influence of 1 variable, the input elements of the damaged and the undamaged configurations differ only by one node. Additionally, since the positive and negative weights have equal strenght, the only possible way for the sum to change sign is if: (a) without damage the node is exactly at the threshold (the Kronecker delta terms in Eqs. \ref{eq:A.3a} and Eq.~\ref{eq:A.3c}), or (b) if after the damage it its exactly at the threshold (the first Kronecker Delta term in Eq.~\ref{eq:A.3b}). Since by damaging a single node we are not able to have the node again at the threshold because of the weights, the term with $h(t)$ in Eq.~\ref{eq:A.3b} cannot be attained. Finally, since $k_d=1$, the only possible values
for $u$, $v$, $w$ and $z=k_d-u-v-w$ are 1 on one of them and 0 on the rest. Using all this information and Eq.~\ref{eq:A.4} we have
\begin{align}
 \mathcal{I} &= \frac{1}{2} \sum_{u=u_0}^{u_f} \sum_{v=v_0}^{v_f} \sum_{w=w_0}^{w_f} Pr(u,v,w) \left\{ H(l-u+w-m+v-z) \delta_{l,m} \right. \nonumber \\
 & \left.+ \delta_{l-u+w,m-v+z}\left[H(m-l)+H(l-m)\right] + H(m-v+z-l+u-w) \delta_{l,m} \right\} \nonumber \\
&= \frac{1}{2} \left\{ \left[\frac{m}{K} + \frac{i-l}{K}\right] \delta_{l,m} + \delta_{l,m-1}\left[\frac{m}{K} + \frac{i-l}{K}\right] \right. \nonumber \\
 & \left.  + \delta_{l-1,m}\left[\frac{l}{K} + \frac{K-i-m}{K}\right] + \left[\frac{l}{K} + \frac{K-i-m}{K}\right] \delta_{l,m} \right\} \nonumber \\
 &= \frac{1}{2K} \left[\delta_{m,l-1}\left(K-i+1\right) + \delta_{m,l+1}\left(i+1\right) \right] + \frac{1}{2}\delta_{l,m} \label{eq:C.4}
\end{align}

\subsubsection*{E.1.1 \quad $p_0\left(b_0\right)$}

To get the result from Eq.~\eqref{eq:C.2} we use the finite Laplace Transform method. Let us define a generating function
\begin{equation}
g(z)=\sum_{k=0}^{\infty} \frac{c_k}{k!} z^k \label{eq:C.5}
\end{equation}
with
\begin{align}
 c_k &= 2^{2k} \left. p_0\left(b_{\infty}\right) \right|_{K=k} = \sum_{i=0}^{k}\binom{k}{i}2^i \ f(k-i), \label{eq:C.6} \\
 f(k-i) &= \sum_{l=0}^{k-i}\binom{k-i}{l} \delta_{l,k-i-l}. \label{eq:C.7}  
\end{align}
Using Eq.~\eqref{eq:C.5} and the definition of $c_k$ given in Eq.~\eqref{eq:C.6}, we obtain
\begin{align}
 g(z)&=\sum_{k=0}^{\infty} \frac{c_k}{k!} z^k \nonumber \\
 &= \sum_{k=0}^{\infty} \sum_{i=0}^{k} \frac{2^i}{i!(k-i)!} z^{k} f(k-i). \nonumber 
\end{align}
With the change of variable $u=k-i$ and inverting the order of summation we get
\begin{align}
 g(z)&= \sum_{k=0}^{\infty} \sum_{i=0}^{k} \frac{\left(2z\right)^i}{i!(k-i)!} z^{k-i} f(k-i) \nonumber \\
 &= \sum_{i=0}^{\infty} \sum_{u=0}^{\infty} \frac{\left(2z\right)^i}{i!u!} z^{u} f(u) \nonumber \\
 &= \left[ \sum_{i=0}^{\infty} \frac{\left(2z\right)^i}{i!} \right] \left[ \sum_{u=0}^{\infty} \frac{z^{u}}{u!}  f(u)\right]. \nonumber
\end{align}
Substituting into this last result the value of $f(u)$ given in Eq.~\eqref{eq:C.7}, making the change of variable $v=u-l$, and again exchanging the sums we get
\begin{align}
 g(z)&= e^{2z} \left[ \sum_{u=0}^{\infty} \sum_{l=0}^{u} \frac{z^{u}}{(u-l)!l!}  \delta_{l,u-l} \right] \nonumber \\
 &= e^{2z} \left[ \sum_{l=0}^{\infty} \sum_{v=0}^{\infty} \frac{z^{2l}}{l!l!} \right] \nonumber \\
 &= e^{2z} I_0(2z),  \label{eq:C.8}  
\end{align}
where $I_0$ is the modified Bessel function of the first kind and where we used the series representation
of the exponential function and $I_0(z)=\sum_{i=0}^{\infty}\left(z/2\right)^{2i}/(i!)^2$.  Using the integral
representation of $I_0(z)$: 
\begin{equation*}
 I_0(z)=\frac{1}{2 \pi }\int_{-\pi}^{\pi} e^{z\cos\theta} d\theta,
\end{equation*}
and the change of variable $\phi=2\theta$ in Eq.~\eqref{eq:C.8} we get
\begin{align}
 g(z)&= \frac{1}{2 \pi }\int_{-\pi}^{\pi} e^{2z\left(\cos\theta+1\right)} d\theta \nonumber \\
   &= \frac{1}{4 \pi }\int_{-2\pi}^{2\pi} e^{4z\cos^2\phi} d\phi \nonumber \\
&= \sum_{k=0}^{\infty} \frac{z^k}{k!} \left[ \frac{4^k}{4 \pi } \int_{-2 \pi }^{2 \pi } \cos^{2k}\phi \ d\phi \right]. \label{eq:C.9} 
\end{align}
Comparing this last result with Eq.~\eqref{eq:C.5} and using the formula
\begin{equation*}
 \int_{0}^{\pi /2} \cos^{2k}\phi \ d\phi=\frac{(2k-1)!!}{(2k)!!}=\frac{1}{2^{2k}}\binom{2k}{k}, 
\end{equation*}
we finally find
\begin{align}
 c_k &= \frac{4^k}{4 \pi } \int_{-2 \pi }^{2 \pi } \cos^{2k}\phi \ d\phi \nonumber \\
 &=  4^k \frac{2}{\pi} \int_{0}^{\pi /2} \cos^{2k}\phi \ d\phi \nonumber \\
 &= \binom{2k}{k} \label{eq:C.10}, 
\end{align}
which, using Eq.~\eqref{eq:C.6}, gives us the first term of Eq.~\eqref{eq:C.1}:
\begin{equation} 
p_0\left(b_0\right)=\frac{1}{2^{2K}}\binom{2K}{K}. \label{eq:C.11} 
\end{equation}

\subsubsection*{E.1.2 \quad $I^{(1)}$}

From Eqs. \eqref{eq:C.3} and \eqref{eq:C.4} we define
\begin{equation}
 I^{(1)}=\frac{1}{2^{2K}} \left(f_1+f_2+f_3\right) \label{eq:C.12},
\end{equation}
where
\begin{align}
 f_1&= \frac{1}{2K} \sum_{i=0}^{K}\binom{K}{i} \ \sum_{l=0}^{i} \sum_{m=0}^{K-i} \binom{i}{l} \binom{K-i}{m} \delta_{m,l-1}\left(K-i+1\right), \label{eq:C.13} \\
 f_2&= \frac{1}{2K} \sum_{i=0}^{K}\binom{K}{i} \ \sum_{l=0}^{i} \sum_{m=0}^{K-i} \binom{i}{l} \binom{K-i}{m} \delta_{m,l+1}\left(i+1\right), \label{eq:C.14} \\
 f_3&= \frac{1}{2} \sum_{i=0}^{K}\binom{K}{i} \ \sum_{l=0}^{i} \sum_{m=0}^{K-i} \binom{i}{l} \binom{K-i}{m} \delta_{m,l}. \label{eq:C.15o} 
\end{align}
To reduce these expressions we will use Vandermonde's identity and the mean value of the hypergeometric function
\begin{align}
 \binom{m+n}{r}&=\sum_{k=0}^r\binom{m}{k}\binom{n}{r-k} \label{eq:C.15} \\
 \sum_{k=0}^r k \binom{m}{k}\binom{n}{r-k} &= \frac{rm}{m+n} \binom{m+n}{r}.\label{eq:C.16}
\end{align}
Using these expressions in $f_1$ with the change of variables $l'=l-1$ and $i'=i-1$
\begin{align}
 f_1&= \frac{1}{2K} \sum_{i=0}^{K}\binom{K}{i} \left(K-i+1\right) \ \sum_{l=0}^{i} \binom{i}{l} \binom{K-i}{l-1} \nonumber \\
 &= \frac{1}{2K} \sum_{i=0}^{K}\binom{K}{i} \left(K-i+1\right) \ \sum_{l'=0}^{i-1} \binom{i}{i-1-l'} \binom{K-i}{l'} \nonumber \\
 &= \frac{1}{2K} \sum_{i=0}^{K} \left(K-i+1\right) \binom{K}{i} \binom{K}{i-1} \nonumber \\
 &= \frac{1}{2K} \sum_{i'=0}^{K-1} \left(K-i'\right) \binom{K}{K-1-i'} \binom{K}{i'} \nonumber \\
 &= \frac{K+1}{4K} \binom{2K}{K-1} = \frac{1}{4}\binom{2K}{K}. \label{eq:C.17}
\end{align}
Doing something similar for $f_2$ with the change of variable $m'=m-1$
\begin{align}
 f_2&= \frac{1}{2K} \sum_{i=0}^{K}\binom{K}{i} \left(i+1\right) \ \sum_{m=0}^{K-i} \binom{i}{m-1} \binom{K-i}{m} \nonumber \\
 &= \frac{1}{2K} \sum_{i=0}^{K}\binom{K}{i} \left(i+1\right) \ \sum_{m'=0}^{K-i-1} \binom{i}{m'} \binom{K-i}{K-i-1-m'} \nonumber \\
 &= \frac{1}{2K} \sum_{i=0}^{K} \left(i+1\right) \binom{K}{i} \binom{K}{K-1-i} \nonumber \\
 &= \frac{1}{2K} \sum_{i=0}^{K-1} \left(i+1\right) \binom{K}{i} \binom{K}{K-1-i} \nonumber \\
 &= \frac{K+1}{4K} \binom{2K}{K-1} = \frac{1}{4}\binom{2K}{K}. \label{eq:C.18}
\end{align}
For $f_3$ we just need to use Eq.~\eqref{eq:C.15} with $m=n=K$
\begin{align}
f_3&= \frac{1}{2} \sum_{i=0}^{K}\binom{K}{i} \ \sum_{l=0}^{i} \binom{i}{l} \binom{K-i}{l}\nonumber \\
&= \frac{1}{2} \sum_{i=0}^{K}\binom{K}{i} \ \sum_{l=0}^{i} \binom{i}{i-l} \binom{K-i}{l}\nonumber \\
&= \frac{1}{2} \sum_{i=0}^{K}\binom{K}{i} \binom{K}{i}  \nonumber \\
&= \frac{1}{2} \sum_{i=0}^{K}\binom{K}{i} \binom{K}{K-i} = \frac{1}{2}\binom{2K}{K}. \label{eq:C.19o}
\end{align}

Using Eqs. \eqref{eq:C.17}, \eqref{eq:C.18} and \eqref{eq:C.19o} in \eqref{eq:C.12} we find the second part of Eq.~\eqref{eq:C.1}:
\begin{equation} 
I^{(1)}=\frac{1}{2^{2K}}\binom{2K}{K}. \label{eq:C.19}
\end{equation}
We finally get the first part of Eq.~\eqref{eq:4.3} using Eqs. \eqref{eq:C.11} and \eqref{eq:C.19} in Eq.~\eqref{eq:C.1}, which gives
\begin{equation} \label{eq:C.20}
  S_0\left(p=0.5,a_G=1,a_R=1,\theta=0\right)=\frac{K+1}{2^{2K}}\binom{2K}{K}.
\end{equation}

\phantomsection
\addcontentsline{toc}{subsubsection}{E.2} 
\subsubsection*{E.2 \quad $S_0$ for $p=0.5$, $a_G=a_R=1$ and $\theta=0.5$} \label{sec:E.2}

In this case the threshold is a noninteger value, $\theta=0.5$. Therefore, from Eq.~\eqref{eq:3.8} and \eqref{eq:3.10}, we have
\begin{equation} \label{eq:C.21}
S_0= KI^{(1)}. 
\end{equation}
where from Eq.~\eqref{eq:3.7c} 
\begin{equation}  \label{eq:C.22}
I^{(1)}= \frac{1}{2^{2K}}\sum_{i=0}^{K}\binom{K}{i} \ \sum_{l=0}^{i} \sum_{m=0}^{K-i} \binom{i}{l} \binom{K-i}{m} \mathcal{I}(a_G=1,a_R=1,\theta=0.5,b=0.5).
\end{equation}
In order to calculate $\mathcal{I}(a_G=1,a_R=1,\theta=0.5,b=0.5)$ we consider the following. The noninteger threshold $\theta=0.5$ makes all the terms with Kronecker deltas effectively zeros, since the exact value of the threshold cannot be attained. Given that $\theta=0.5>0$, it also makes nodes in which the sum of the updating function equals $0$, $\sum_{j}a_{n,j}\sigma_{n_j}=0$, become inactive, as this sum is smaller than the threshold. As a consequence, the only way in which the changing of one of the inputs of a node changes the state of the target node is if either the sum before the damage was $0$ and after the damage it is above $0$, or vice versa. In addition, since $k_d=1$, then the only possible values for $u$, $v$, $w$ and $z=k_d-u-v-w$ are 1 on one of them and 0 on the rest. Using these facts and Eq~\eqref{eq:A.4} we get
\begin{align}
 \mathcal{I} &= \sum_{u=u_0}^{u_f} \sum_{v=v_0}^{v_f} \sum_{w=w_0}^{w_f} Pr(u,v,w) \left\{ H(l-u+w-m+v-z-0.5) \right. \nonumber \\
& \left. \times H(m-l+0.5) + H(m-v+z-l+u-w+0.5) H(l-m-0.5) \right\} \nonumber \\
&= \left[\frac{m}{K} + \frac{i-l}{K} \right] \delta_{l,m} + \left[\frac{l}{K} + \frac{K-i-m}{K} \right] \delta_{l,m+1} \nonumber \\
&= \frac{K-i+1}{K}~\delta_{l,m+1} + \frac{i}{K}~\delta_{l,m}. \label{eq:C.23}
\end{align}
Using this result in Eq.~\eqref{eq:C.22} we have
\begin{equation}  \label{eq:C.24}
I^{(1)}= \frac{1}{2^{2K}} \left(g_1 + g_2\right),
\end{equation}
where
\begin{align}
 g_1&= \frac{1}{K} \sum_{i=0}^{K}\binom{K}{i} \ \sum_{l=0}^{i} \sum_{m=0}^{K-i} \binom{i}{l} \binom{K-i}{m} \delta_{l-1,m}\left(K-i+1\right), \label{eq:C.25} \\
 g_2&= \frac{1}{K} \sum_{i=0}^{K}\binom{K}{i} \ \sum_{l=0}^{i} \sum_{m=0}^{K-i} \binom{i}{l} \binom{K-i}{m}~i \delta_{l,m}. \label{eq:C.26}
\end{align}
From Eqs. \eqref{eq:C.13} and \eqref{eq:C.25} it follows that
\begin{equation} \label{eq:C.27}
g_1=2f_1=\frac{1}{2}\binom{2K}{K},
\end{equation}
Thus, we only need to calculate $g_2$. By using Vandermonde's identity and the mean value of the hypergeometric function, Eqs. \eqref{eq:C.15} and \eqref{eq:C.16}, we obtain
\begin{align}
g_2&= \frac{1}{K} \sum_{i=0}^{K}\binom{K}{i}~i \ \sum_{l=0}^{i} \binom{i}{l} \binom{K-i}{l}\nonumber \\
&= \frac{1}{K} \sum_{i=0}^{K}\binom{K}{i}~i \ \sum_{l=0}^{i} \binom{i}{i-l} \binom{K-i}{l}\nonumber \\
&= \frac{1}{K} \sum_{i=0}^{K}\binom{K}{i} \binom{K}{i}~i \nonumber \\
&= \frac{1}{K} \sum_{i=0}^{K}\binom{K}{i} \binom{K}{K-i}~i = \frac{1}{K} \frac{K}{2} \binom{2K}{K} =\frac{1}{2}\binom{2K}{K}. \label{eq:C.28}
\end{align}
Using Eqs. \eqref{eq:C.24}, \eqref{eq:C.27} and \eqref{eq:C.28} in Eq.~\eqref{eq:C.21}, we get the desired result (the $\theta=0.5$ case of Eq.~\eqref{eq:4.3}):
\begin{equation} \label{eq:C.29}
  S_0\left[p=0.5,a_G=1,a_R=-1,\theta=0.5\right]=\frac{K}{2^{2K}}\binom{2K}{K}.
\end{equation}

\phantomsection
\addcontentsline{toc}{subsubsection}{E.3} 
\subsubsection*{E.3 \quad $S_0\left(p=0.5,a_G=1,a_R=1,\theta=-0.5\right)$} \label{sec:E.3}

Since, $\theta=-0.5$, from Eqs. \eqref{eq:3.8}, \eqref{eq:3.10} and \eqref{eq:3.7c} we get
\begin{align} 
S_0&= KI^{(1)}, \label{eq:C.30} \\ 
I^{(1)}&= \frac{1}{2^{2K}}\sum_{i=0}^{K}\binom{K}{i} \ \sum_{l=0}^{i} \sum_{m=0}^{K-i} \binom{i}{l} \binom{K-i}{m} \mathcal{I}(a_G=1,a_R=1,\theta=-0.5,b=0.5).
\end{align}
To calculate $\mathcal{I}(a_G=1,a_R=1,\theta=-0.5,b=0.5)$ we look back at the derivation of $\mathcal{I}(a_G=1,a_R=1,\theta=0,b=0.5)$ in the last section (Sec. \ref{sec:E.2}) and notice that both cases are similar. The difference lies in that, in this case, the threshold $\theta=-0.5<0$. This threshold then makes the nodes in which the sum of the updating function equals $0$, (namely for which $\sum_{j}a_{n,j}\sigma_{n_j}=0$), become active, as the sum is larger than the threshold. Therefore, the only cases in which damaging one of the inputs of a node changes the state of the target node is if either the sum before the damage was $0$ and after damage it is below $0$, or vice versa. Again, since $k_d=1$, the only possible values for $u$, $v$, $w$ and $z=k_d-u-v-w$ are 1 on one of them and 0 on the rest. Using all this in Eq.~\eqref{eq:A.4} we obtain
\begin{align}
 \mathcal{I} &= \sum_{u=u_0}^{u_f} \sum_{v=v_0}^{v_f} \sum_{w=w_0}^{w_f} Pr(u,v,w) \left\{ H(l-u+w-m+v-z+0.5) \right. \nonumber \\
& \left. \times H(m-l-0.5) + H(m-v+z-l+u-w-0.5) H(l-m+0.5) \right\} \nonumber \\
&= \left[\frac{m}{K} + \frac{i-l}{K} \right] \delta_{l+1,m} + \left[\frac{l}{K} + \frac{K-i-m}{K} \right] \delta_{l,m} \nonumber \\
&= \frac{i+1}{K}~\delta_{l+1,m} + \delta_{l,m} - \frac{i}{K}~\delta_{l,m}. \label{eq:C.32}
\end{align}
Using this result in Eq.~\eqref{eq:C.22} we get
\begin{equation}  \label{eq:C.33}
I^{(1)}= \frac{1}{2^{2K}} \left(h_1 + h_2 - h_3\right),
\end{equation}
where
\begin{align}
 h_1&= \frac{1}{K} \sum_{i=0}^{K}\binom{K}{i} \ \sum_{l=0}^{i} \sum_{m=0}^{K-i} \binom{i}{l} \binom{K-i}{m} \delta_{l+1,m}\left(i+1\right), \label{eq:C.34} \\
 h_2&= \sum_{i=0}^{K}\binom{K}{i} \ \sum_{l=0}^{i} \sum_{m=0}^{K-i} \binom{i}{l} \binom{K-i}{m} \delta_{l,m}. \label{eq:C.35} \\
 h_3&= \frac{1}{K} \sum_{i=0}^{K}\binom{K}{i} \ \sum_{l=0}^{i} \sum_{m=0}^{K-i} \binom{i}{l} \binom{K-i}{m}~i \delta_{l,m}. \label{eq:C.36}
\end{align}
By comparing Eq.~\eqref{eq:C.14} with \eqref{eq:C.34} and Eq.~\eqref{eq:C.26} with \eqref{eq:C.36}, we have
\begin{align} 
h_1&=2f_2=\frac{1}{2}\binom{2K}{K}, \label{eq:C.37} \\
h_3&=g_2=\frac{1}{2}\binom{2K}{K}. \label{eq:C.38}
\end{align}
The only term that has not been calculated yet is $h_2$. To do this we use Vandermonde's identity, Eq.~\eqref{eq:C.15}, with $m=n=K$
\begin{align}
h_2&= \sum_{i=0}^{K}\binom{K}{i} \ \sum_{l=0}^{i} \binom{i}{l} \binom{K-i}{l}\nonumber \\
&= \sum_{i=0}^{K}\binom{K}{i} \ \sum_{l=0}^{i} \binom{i}{i-l} \binom{K-i}{l}\nonumber \\
&= \sum_{i=0}^{K}\binom{K}{i} \binom{K}{i} \nonumber \\
&= \binom{2K}{K}. \label{eq:C.39}
\end{align}
Finally we get the $\theta=-0.5$ case of Eq.~\eqref{eq:4.3} using Eqs. \eqref{eq:C.37}, \eqref{eq:C.38}, \eqref{eq:C.39} and \eqref{eq:C.33} in Eq.~\eqref{eq:C.30},
\begin{equation} \label{eq:C.40}
  S_0\left(p=0.5,a_G=1,a_R=1,\theta=-0.5\right)=\frac{K}{2^{2K}}\binom{2K}{K},
\end{equation}
which is, of course, the same as the case  $\theta=0.5$ because of the symmetry of the weights and the probabilities.

%
%

\end{document}